# A Boltzmann statistical approach for the analysis of polarization states in mixed phase ferroelectric materials: application to morphological phase boundary

A. Pramanick[1,2], L. Daniel[1,2]

[1]*Université Paris-Saclay, CentraleSupélec, CNRS, Laboratoire de Génie Électrique et Électronique de Paris, 91192, Gif-sur-Yvette, France*

[2]*Sorbonne Université, CNRS, Laboratoire de Génie Électrique et Électronique de Paris, 75252, Paris, France*

abhijit.pramanick@centralesupelec.fr

laurent.daniel@centralesupelec.fr

# Abstract

Ferroelectrics are widely used for a broad array of technological applications due to their attractive electrical and electromechanical properties. In order to obtain large functional properties, material compositions are often designed to favor a coexistence of multiple ferroelectric phases. For such compositions, the macroscopically observed properties are variously attributed to easier domain switching and/or phase transition. Nevertheless, modelling of concurrent domain switching and phase transition in mixed phase ferroelectrics remains a challenging task. Here, a methodology is presented to quantitatively evaluate the volume fractions of different domain variants in a mixed phase ferroelectric under complex electromechanical loading. The methodology combines the phenomenology of Landau free energy of ferroelectric phases with Boltzmann statistical analysis, and is presented for Pb(Zr,Ti)O$_3$ near morphotropic phase boundary (MPB). It is shown that specific grain orientation has a significant effect on how proximity to phase boundary affects microscopic phenomena at the single-crystal level. An estimate of phase and domain switching behavior in a polycrystalline aggregate is subsequently obtained, and the resultant polarization and strain responses at the macroscopic level are computed for a material with random texture. The results indicate the progressive evolution of domain and phase fractions in a material near MPB with mixed ferroelectric phases. We show that in polycrystalline materials with compositions slightly on the tetragonal side of MPB, grains that exhibit large 90° domain switching have a larger contribution to the macroscopic strain response as compared to grains that undergo tetragonal-to-rhombohedral phase-switching.

# 1. Introduction

Ferroelectrics are widely used in many modern technologies including high-power capacitors, memories, precision actuators and sensors, MEMS, pyroelectric energy harvesters and electrocaloric cooling.[1-4] In order to undertake precise design of engineering components for these applications, it is important to develop appropriate modelling tools for predicting the performance of ferroelectric materials under different operating conditions. Although considerable works have been undertaken towards this end in the past, improvement over past approaches is desired for various reasons, as follows. First, as the scope of applications of ferroelectric materials broadens, they are subjected to more extreme conditions, including higher electric-fields, multiaxial large stress fields, or wider temperature variations.[5-9] Second, with increased complexity of engineering components that is facilitated by recent developments in additive manufacturing technologies, the different parts of even the same material component can experience widely different external-stimuli, for example stress or electric-field concentration at nodes in an architected ferroelectric component.[10,11] The combination of microscopic mechanisms over such extreme conditions can deviate from the more well-known mechanisms based on which the current models are proposed. Third, driven by a desire to achieve higher engineering performance, ferroelectric compositions are often placed in close proximity to phase boundaries for both Pb-based $Pb(Zr,Ti)O_3$ (PZT) as well as Pb-free ferroelectrics, such as solid solutions of $BaTiO_3$, $(K,Na)NbO_3$ and $(Na,Bi)TiO_3$.[12,13] The microscopic mechanisms operational near phase boundaries can be diverse, and may also change with small changes in composition or temperature, which has not been explored fully. All of these above factors necessitate the development of new modelling tools.

The functional properties of ferroelectrics fundamentally arise from their spontaneous electrical polarization, whose state can be changed by the application of an electric field, stress or temperature variations.[14-16] The development of spontaneous polarization in a ferroelectric material below its Curie temperature ($T_C$) is, in turn, accompanied by its transformation from a high-temperature cubic phase to one or more non-centrosymmetric polar crystal phase(s) at temperatures below $T_C$. For temperatures lower than $T_C$, the material can exist in either single or multiple ferroelectric phases as defined by their crystallography symmetries, such as tetragonal, orthorhombic or rhombohedral. For each of these phases, there can exist multiple domain variants in which the polar vectors are misoriented by either 180° or non-180° depending on specific crystallography symmetry. For example, the polarization vectors in adjacent domains can be misoriented by 90° in tetragonal phase, 71°/109° in rhombohedral phase, in addition to possible misorientation of 180° for all two phases. Under the application of an external stimulus, such as electric-field or stress, the polarization vectors can be reoriented in a reversible or irreversible manner into directions that are allowed by crystallography symmetry, which is defined as domain switching. In cases where multiple phases co-exist, or where it is energetically favorable, application of an external stimuli may also trigger a broader realignment of polarization vectors through phase transformation. The contributions from domain switching and phase transformation to the functional response of a ferroelectric is generally categorized as the *extrinsic response*. In addition, part of the macroscopic response of a ferroelectric also comes from lattice strain caused by piezoelectric effect, which is categorized as the *intrinsic response.* In reality, multiple of these mechanisms can simultaneously occur, and all of them should ideally be considered while evaluating the macroscopic response of a ferroelectric material. However, for computational ease, in some cases only the major contributions may be considered to obtain approximate solutions.

From a thermodynamic viewpoint, the overall behavior of a ferroelectric can be divided into a rate-independent reversible component and a rate-dependent irreversible component.[16] Different modelling approaches have been developed to model these components independently. Within the framework of Landau-Ginzburg-Devonshire (LGD) theory, the rate-independent reversible behavior of a single-crystal ferroelectric can be predicted from thermodynamic models in which the Gibbs free-energy of a material is expressed as a Taylor series expansion of appropriate order parameters, such as spontaneous polarization or spontaneous strain.[17-24] In principle, minimization of the free energy function can then be used to predict the stability of a particular phase or possible distortions within a single phase under different conditions of temperature, stress or electric-field. Note that minimization of the free energy function can result in multiple solutions for minima within the three dimensional orientation space, which can be interpreted as different polar domains within a single-crystal material. In general, these models have not been used to predict the evolution of the volume fractions of the different polar states in a material.

In contrast to lattice distortion/phase transformation, domain switching is considered as the major contributor to the rate-dependent irreversible component of the total response of a ferroelectric material. To deal with the rate-dependent irreversible component, mainly two different approaches have been developed. In one approach, the switching between the different domain variants is predicted based on the rate of dissipation of the Helmholtz free energy term due to changes in remanent strain and polarization states.[25-29] This approach is based on irreversible thermodynamic models that were originally developed to describe crystal plasticity. A switching criterion can then be developed, which requires that this energy dissipation under external driving forces must be non-negative. An alternate approach is to use phase-field models in which rate-dependent Landau equations is applied to model the evolution of domain and phase microstructures.[30-32] Both of these approaches can be computationally intensive, especially if one considers the possibility of a wide range of polarization states, such as is the case for co-existing ferroelectric phases. In order to expedite the computation process, Daniel *et al.* proposed an alternative approach in which the population of the different domain variants is calculated based on Boltzmann statistical distribution.[33] Inspired by similar formulations for ferromagnetic materials,[34,35] Daniel *et al.* proposed that the relative volume fractions of the different domain variants are related to their Helmholtz free energies, that is, the useful work performed by applied electric-field and/or stress.[35] This method allows for calculation of the anhysteretic reversible changes in domain populations under applied electric-field or stress. The rate-dependent irreversible part to the domain switching fraction can then be calculated by secondary phenomenological approaches, such as refs.[36-39].

    The method described in ref.[33] is appealing because of its simplicity and computational ease; however, it requires modification for application to more complex scenarios, such as co-existing multiple phases, temperature changes or induced phase transitions. This is because, the Helmholtz free energy terms used in the formulation for Boltzmann statistical distribution in that work does not incorporate effect of temperature changes or free energy changes due to phase transition. In order to take into effect these later aspects, we propose here an extension of the approach described in ref.[33-35]. Specifically, we propose incorporation of the Landau free energy for application of Boltzmann statistical law for computation of the

volume fractions of the different domain or phase variants. The use of the Landau free energy, instead of only the Helmholtz free energy terms, broadens the applicability of the model proposed in ref. [33] to more complex scenarios for actual materials in service, which often encounter electric-fields or stress in combination with temperature changes and can experience multiple microscopic phenomena including domain switching and/or phase transitions.

The objective here is to quantitatively evaluate the volume fractions of domain variants with different polarization orientations in a material with mixed ferroelectric phases. We propose a methodology that combines the phenomenology of Landau free energy of ferroelectric phases with Boltzmann statistical analysis. We present application of the methodology for the case of coexisting tetragonal and rhombohedral ferroelectric phases near a morphotropic phase boundary (MPB) of the solid-solution of $Pb(Zr,Ti)O_3$. The methodology presented is generic and has the potential to be applied for more complex systems, such as more than two coexisting phases or even highly disordered systems such as relaxors. First, we lay down the basic phenomenology for application of Landau free energy within the framework of Boltzmann statistical distribution to predict reversible changes in domain and phase fractions in a ferroelectric material near morphotropic phase boundary (MPB) at the single-crystal level. Second, we apply the methodology to predict the collective behavior of a polycrystalline aggregate with random distribution. The modelling results are compared with reported experimental measurements of domain and phase switching in PZT ceramics near MPB composition.

## 2. Basic phenomenology

We first examine how the proportion of the various possible polar distortions changes while traversing across a phase transition point, as function of composition. We start with the thermodynamic expression of the Gibbs free energy, $\Delta G$, in the framework of the Landau-Ginsburg-Devonshire (LGD) phenomenological model, under zero electric-field and zero stress, and expanded to order 6:[17-24]

$$\Delta G = \alpha_1[P_1^2 + P_2^2 + P_3^2] + \alpha_{11}[P_1^4 + P_2^4 + P_3^4] + \alpha_{12}[P_1^2 P_2^2 + P_2^2 P_3^2 + P_3^2 P_1^2] +$$
$$\alpha_{111}[P_1^6 + P_2^6 + P_3^6] + \alpha_{112}[P_1^4(P_2^2+P_3^2) + P_2^4(P_3^2+P_1^2) + P_3^4(P_1^2+P_2^2)] + \alpha_{123}P_1^2 P_2^2 P_3^2 \quad (1)$$

where $P_1, P_2$ and $P_3$ refer to the components of the polarization **P** of a ferroelectric phase along the three axes of the coordinate system of its parent cubic paraelectric phase. In eq.(1), the coefficients $\alpha$ refer to the respective dielectric stiffnesses at constant dielectric displacement and constant stress. The stability of the various ferroelectric phases can be determined by equating $\frac{\partial \Delta G}{\partial P_i} = 0$, while applying specific symmetry conditions for the different polarization components; for example, $P_1 = P_2 = 0$ and $P_3 \neq 0$ for the tetragonal phase, $P_1 = 0$ and $P_2 = P_3 \neq 0$ for the orthorhombic phase and so on. Also, in eq.(1), $\alpha_1$ is assumed to be temperature dependent, while all other coefficients are considered to be temperature independent.

The LGD phenomenological theory has been successfully used to describe composition dependent phase transition across a morphological phase boundary (MPB), such as in lead zirconate titanate Pb(Zr,Ti)O$_3$,[19-24] as well as temperature-dependent phase transition across a polymorphic phase boundary, such as in BaTiO$_3$. [40,41] However, in many practical cases, one encounters a coexistence of different phases and therefore an important objective is to determine how the relative volume fractions of domains with different polar states change under applied electric fields and/or stress - this is the problem we address below.

## 3. Computation at the single-crystal level

We can apply eq.(1) to evaluate the three dimensional Landau energy surface for different compositions of Pb(Zr,Ti)O$_3$. The schematic in Figure 1(a) illustrates the possible deviation of the polarization vector from the 001 axis, which is defined in terms of the angles $\psi$ and $\theta$, where $\psi$ represents the rotation angle from the 100 axis and $\theta$ represents the azimuthal angle from the 001 axis. The free energy for each orientation of the polarization vector is calculated following equation (1). The values of the coefficients $\alpha$ are obtained from Haun *et al.*[21-23] (see table in Appendix 3). Figure 1(b) shows the profile of the Landau free energy $G$ (in J/m$^3$) as a function of different Zr/Ti ratios in the Pb(Zr,Ti)O$_3$ solid-solution for the (010) plane, that is $\phi$ = 0°. For this plane, the minima in G are observed at $\theta$ = 0°, $\theta$ = $\pm$90° and $\theta$ = $\pm$180°, which represent the orientations of the polarization vector within the different domain variants of the tetragonal phase. Note that the minima become sharper with increasing Ti ratio, which indicates that the tetragonal polar states become highly stable against possible deviations for higher Ti content. Figure 1(c) shows the profile of the Landau free energy G as a function of different Zr/Ti ratios in the Pb(Zr,Ti)O$_3$ solid-solution for the $(1\bar{1}0)$ plane, that is $\phi$ = 45°. For the $(1\bar{1}0)$ plane, the minima in G are observed at $\theta$ = 0° and $\theta$ = $\pm$180°, which represent the 180° domain variants of tetragonal-like polar states, as well as at $\theta$ = $\pm$54.7° and $\theta$ = $\pm$125.3°, which represents orientation of the polarization vectors within the different domain variants of rhombohedral-like polar states that are oriented along the <111> crystallographic directions. The energy for the minima with tetragonal and rhombohedral-like polar distortions have the same value for Zr/Ti ratio of 52/48; however, the minima for the rhombohedral polar distortions increase relative to the tetragonal polar distortions with increasing Ti content. For intermediate values of ϕ between 0° and 45°, the free energy profile shows a gradual transition between these two extremes. Figure 1(d) and (e) show the complete energy surfaces for Pb(Zr,Ti)O$_3$ with reference to the cartesian coordinate system of the parent cubic phase for compositions of Zr/Ti ratio of 52/48 and 46/54, respectively.

In order to determine the probability of the various orientations of the polarization vector, we introduce an internal variable at the single-crystal level, $f_a$. Here, $f_a$ represents the volume fraction of a family of domains, which have the same orientation of the polarization vector that is defined by *a*. The value of $f_a$ under equilibrium condition can be explicitly calculated following the Boltzmann probability function, equation (2), such as described earlier for both ferromagnetic and ferroelectric systems.[33-35]

$$f_a = \frac{\exp(-A.G_{i=a})}{\sum_{i=1}^{K} \exp(-A.G_i)} \qquad (2)$$

In equation (2), $G_i$ is the free energy for a probable orientation of the polarization vector, which is based on one of the *K* minima in the free energy surface. *A* is an adjustment parameter, which can be related to the range of compositions over which the phase transition occurs as well as the slope of the dielectric susceptibility of a particular phase near zero electric-field, as we will see below. In equation (2), the denominator is obtained by summation of the factor $\exp(-A.G_i)$ over all probable orientations of polarization vector that are defined by the minima in the free energy surface.

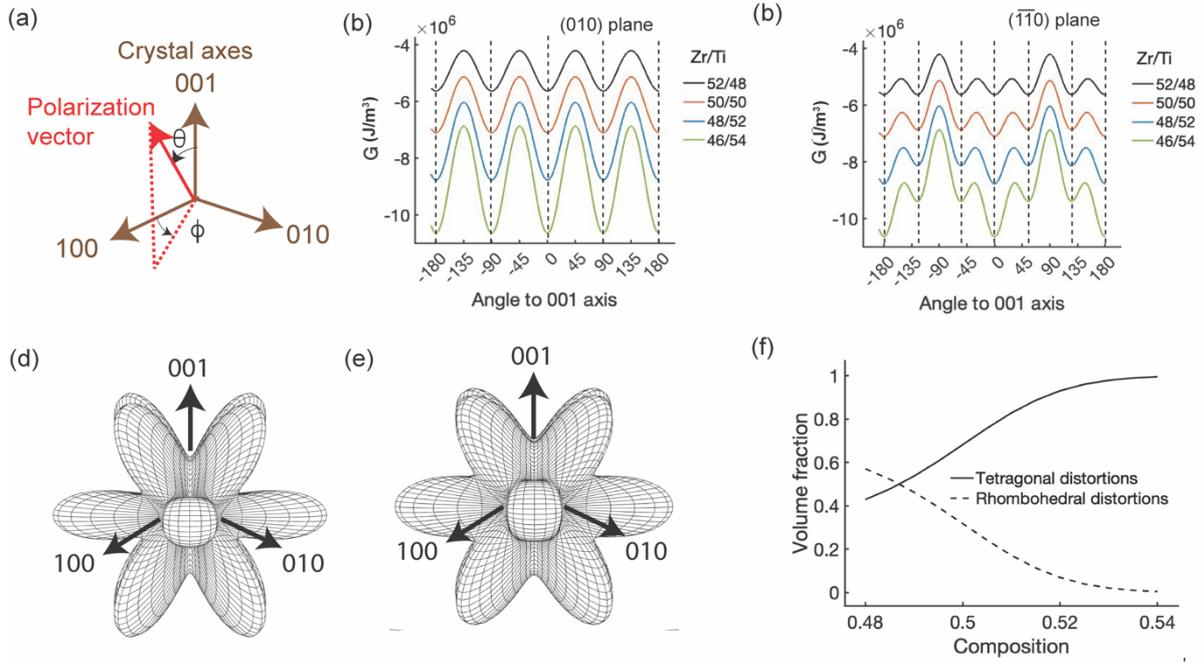

**Figure 1:** (a) Illustration of modelling set-up for calculation of Landau free energy, G, corresponding to certain orientation of polarization vector, which is defined by the two angles ϕ and θ. (b) G for polarization vector orientations within the $(0\bar{1}0)$ crystallographic plane, as function of angle to the 001 axis, θ, and for various Zr/Ti ratio in Pb(Zr,Ti)O$_3$. (c) same as in (b), for $(1\bar{1}0)$ crystallographic plane. (d,e) Landau free energy surface for Zr/Ti ratio of 52/48 and 46/54, respectively. (f) Volume fraction of polar states corresponding to tetragonal and rhombohedral distortions, as a function of Ti content

For example, for ϕ = 45°, there are six minima (counting one of either +180° or -180° orientations), which have the same value of $G$ for the composition Zr/Ti =52/48. Hypothetically, if all the polarization vectors within the single-crystal are to reside within the crystallographic plane defined by ϕ = 45°, then application of equation (2) predicts equal volume fractions for each of the six orientations of the polarization vector. Considering the polarization vectors with θ = ±54.7° and θ = ±125.3° to belong to rhombohedral-like states and θ = 0° and θ = ±180° to belong to the tetragonal-like states, the relative volume fractions of the rhombohedral and tetragonal phases are 2/3 and 1/3, respectively. However, in order to calculate the relative proportions of all the probable orientations of the polarization vector, we must identify all the local minima from the three-dimensional energy surface, such as shown in Figure 1(d,e), and subsequently apply eq.(2) over all the probable polar states. For the Zr/Ti=52/48 composition, there are six possible tetragonal-like polar distortions and eight possible rhombohedral-like polar distortions, all of which have the same minimum energy value. Accordingly, application of eq.(2) predicts the volume fraction of tetragonal domain variants to be equal to 6/14 and that for tetragonal domain variants to be equal to 8/14. With increasing Ti content, the minima for the rhombohedral states are higher as compared to that of the tetragonal states, and accordingly as per equation (2), the relative volume fraction of rhombohedral domains will decrease. The rate of change in volume fractions of the tetragonal and rhombohedral states as function of composition depends on the value of $A$. The values of fractions of tetragonal and rhombohedral domain variants as a function of composition, calculated for $A$ = 1e-5 m$^3$/J, near the MPB are shown in Figure 1(f). The relative volume fraction of rhombohedral variants asymptotically approaches zero for Zr/Ti ratio ~ 46/54.

3.1 Application of electric -field

As shown by Damjanovic,[16] the LGD phenomenological model can be used to predict the susceptibility of polar distortions from the ideal symmetry of a specific ferroelectric phase under small disturbances. For example, in the tetragonal phase, one can characterize the susceptibility for rotation of the polarization vector from the ideal polar 001 axis by evaluating the complete energy surface in the cubic coordinate system and considering the flatness of the energy surface in the three dimensional orientation space. Here, we explore the use of Boltzmann statistics to predict the change in fractions of different polar states under the application of an electric-field.

In order to identify the possible polar distortions under the application of an external electric field, eq.(1) can be further expanded as follows:

$$\Delta G = \alpha_1[P_1^2 + P_2^2 + P_3^2] + \alpha_{11}[P_1^4 + P_2^4 + P_3^4] + \alpha_{12}[P_1^2 P_2^2 + P_2^2 P_3^2 + P_3^2 P_1^2] +$$
$$\alpha_{111}[P_1^6 + P_2^6 + P_3^6] + \alpha_{112}[P_1^4(P_2^2+P_3^2) + P_2^4(P_3^2+P_1^2) + P_3^4(P_1^2+P_2^2)] + \alpha_{123} P_1^2 P_2^2 P_3^2 -$$
$$P_1 E_1 - P_2 E_2 - P_3 E_3 \qquad (2)$$

where $E_i$ refer to the electric-field components along the three axes of the coordinate system of the parent cubic paraelectric phase.

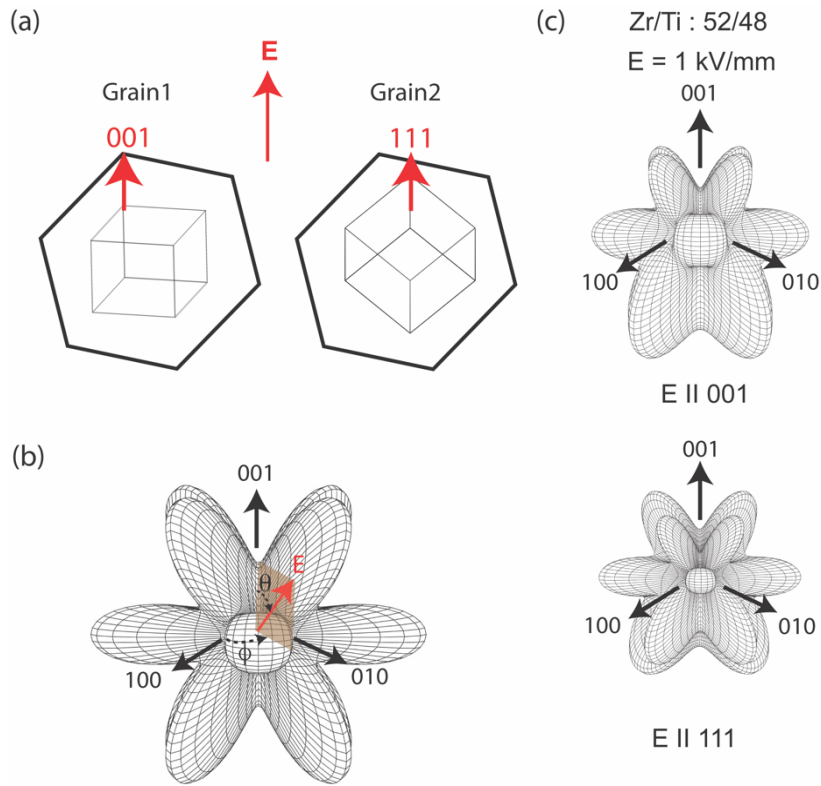

**Figure 2:** (a) Illustrative example of orientation of two different grains with respect to applied electric field. (b) Definition of electric field direction with respect to crystallographic directions, as defined by the angles $\phi_E$ and $\theta_E$. (c) Illustration of distortion of Landau free energy surface for two different directions of applied electric field, **E**.

Note that in a polycrystalline material, the orientation of the electric-field for each single crystallite will be different depending on its relative orientation with respect to the macroscopic electric-field direction. This is illustrated in Figure 2(a).

For Grain 1, the electric-field is applied parallel to the 001 crystallographic axis, while for Grain 2, the electric-field is applied parallel to the 111 crystallographic axis. In order to determine the responses of these two different grains, we will need to evaluate the components $E_1$, $E_2$ and $E_3$ of the electric field $\boldsymbol{E}$ along the three axes of the cartesian coordinates of the parent cubic phase. To generalize the situation, we can determine the distortion of the Landau energy surface for different orientations of electric-field, as illustrated in Figure 2(b), where ϕ_E and θ_E define the orientation of the electric-field direction with respect to the cubic coordinate axes. Figure 2(c) shows the distortion of the Landau energy surface for the composition 52/48, and for $\boldsymbol{E}$ ∥ 001 and $\boldsymbol{E}$ ∥ 111. The relative distortions of the energy surfaces in these two situations influence the minima in energy corresponding to the different polar distortions, and consequently, the relative volume fractions of the tetragonal and rhombohedral domain variants. The degree to which the energy surface is distorted under an applied electric-field also depends on the relative value of the electric-field energy term $-\boldsymbol{P}.\boldsymbol{E}$ with respect to the minima in the energy surface under zero field condition.

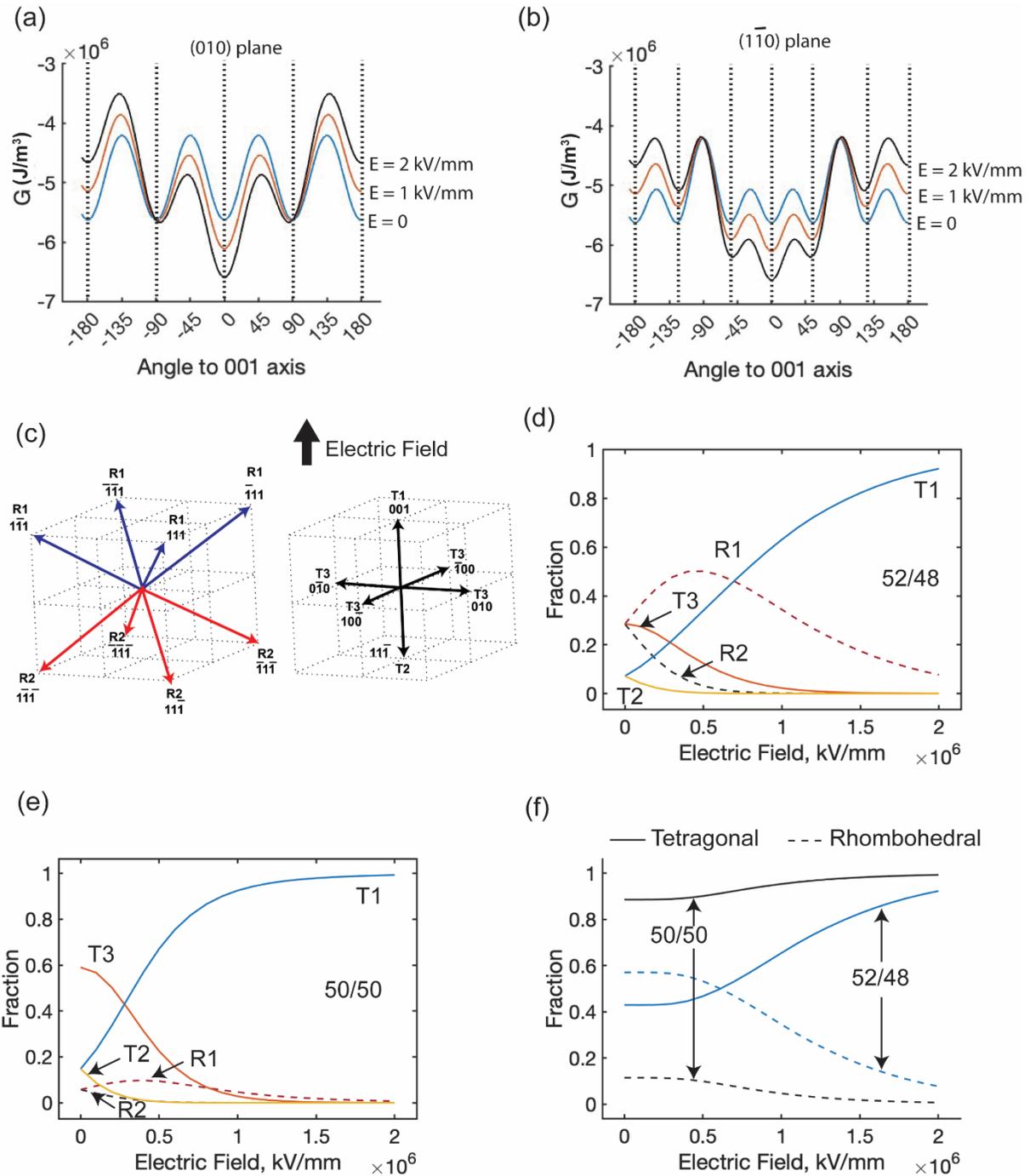

**Figure 3:** (a,b) Line profiles of $G$ for polarization vectors residing within (010) and ($1\bar{1}0$) atomic planes, respectively, and for different electric-field magnitudes applied parallel to 001 direction. (c) Type of polar domains as defined by the orientation of their polarization vectors with respect to the electric field direction. For example, domains marked R1 have their polarization vector oriented 54.7° with respect to electric-field direction. (d,e) Change in volume fraction of different domain variants as function of applied electric-field magnitude for two different values of Zr/Ti ratio. (f) Change in overall volume fractions of tetragonal and rhombohedral domains as function of applied electric field and for two different values of Zr/Ti ratio, for the specific orientation of the grain with respect to the electric field shown in fig. 3(c).

Figure 3(a) shows $G$ for polarization vectors within the plane (010) and with orientation θ with respect to the 001 axis, for three different values of electric-field applied parallel to the 001 direction (as illustrated for grain 1 in Figure 2(a)). It is clear that with increasing electric-field magnitude, the domains with polarization vectors parallel to 001 have a lower free energy as compared to those with polarization vector having orientations of θ=±90° and θ=±180°. For polarization within the $(1\bar{1}0)$ plane (as illustrated for grain 2 in Figure 2(a)), Figure 3(b) similarly shows changes in the energy-levels of the different possible polar states under applied electric fields. For the latter case, the energy level corresponding to the polarization vector parallel to 001 (that is, θ=0°) is lower than that corresponding to orientations of θ equal to ±54.7°, ±125.3° and ±180°. However, the difference in energy levels corresponding to θ=0° and θ=±54.7° is lower than the same calculated for θ=0° and θ=±90°. Accordingly, depending on changes in their relative energy levels, the volume fractions of different domain variants as function of electric-field magnitude can be calculated using eq. (2). Figure 3(c) illustrates the different possible orientations of the polarization vectors within the tetragonal and rhombohedral domain variants, and their relation to the applied electric field direction. The polar domains are defined as per the orientation of their polarization vectors with respect to the electric field direction. For example, domains marked R1 and R2 have their polarization vector oriented ±54.7° and ±125.3° with respect to electric-field direction, respectively. Similarly, domains marked T1 have their polarization parallel to the applied electric field direction, while those marked T2 and T3 have their polarization vector oriented 180° and ±90° with respect to electric-field direction, respectively.

Figure 3(d) shows changes in volume fractions of these different domain variants with increasing electric field. For the tetragonal domain variants, polarization vector for T1 is favorably oriented with respect to the electric-field direction, and therefore domains with this polarization orientation grow in volume fraction with increasing electric field at the expense of domains T2, T3 and R2. Interestingly, volume fraction of domains marked R1 initially increases with increasing electric-field, but subsequently decreases for higher electric-field magnitudes. In physical terms, this can be interpreted as a process whereby switching between 90° domain variants of the tetragonal phase proceeds through an intermediate transition from unfavorably oriented tetragonal domains (T2 and T3) to favorably oriented rhombohedral domains (R1). For electric-field magnitudes higher than 1 kV/mm, the polarization process is solely determined by transition from R1 to T1 domain variants.

Figure 3(e) presents changes in different domain variants under the same set of conditions as described above, but for the Zr/Ti ratio of 50/50. In this case, the overall volume fraction of rhombohedral phase variants is much lower as compared to that for the 52/48 (Zr/Ti) composition. Consequently, as can be seen from Figure 3(e), the dominant process for polarization switching in this case is the direct transition from T3 domains to T1 domains, that is 90° domain switching.

Figure 3(f) shows the relative volume fraction of the tetragonal and rhombohedral phase variants as function of electric field for two different compositions. It is clear that phase switching plays an important part for electric-field induce polarization change in the 52/48 composition, but almost no role for the 50/50 composition. In other words, the predominance of a specific mechanism towards polarization switching, viz. switching between domains of a ferroelectric phase or switching between different ferroelectric phases, is dependent of the relative stability of these polar states as determined by Boltzmann statistics.

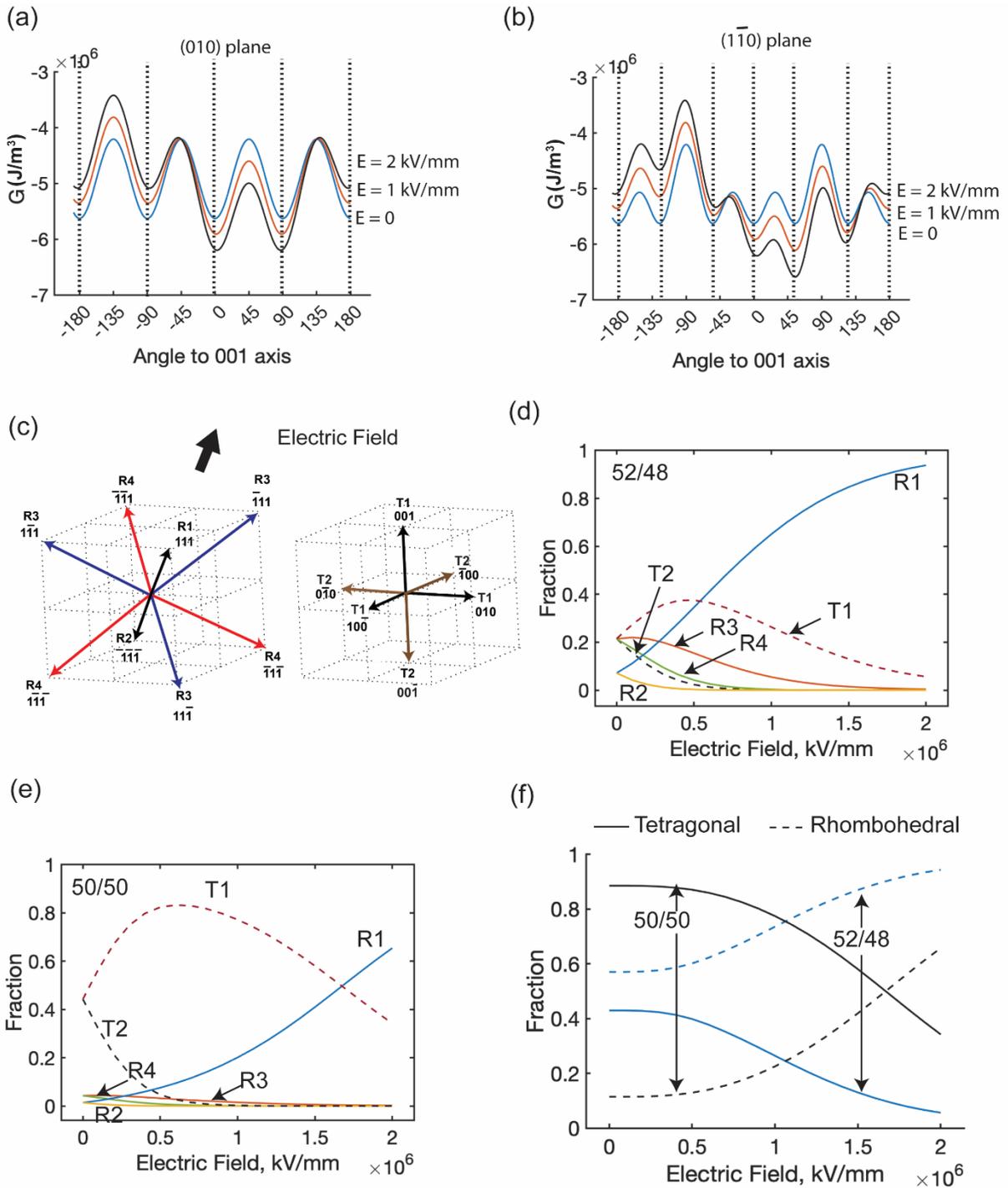

**Figure 4:** (a,b) Line profiles of $G$ for polarization vectors residing within (010) and ($1\bar{1}0$) atomic planes, respectively, and for different electric-field magnitudes applied parallel to 111 direction. (c) Type of polar domains as defined by the orientation of their polarization vectors with respect to the electric field direction. For example, domains marked T1 have their polarization vector oriented 54.7° with respect to electric-field direction. (d,e) Change in volume fraction of different domain variants as function of applied electric-field magnitude for two different values of Zr/Ti ratio. (f) Change in overall volume fractions of tetragonal and rhombohedral domains as function of applied electric field and for two different values of Zr/Ti ratio.

Figure 4 presents results for changes in different domain variants of the tetragonal and rhombohedral phases for electric-field applied along the 111 direction, that is, for the situation illustrated for Grain 2 in Figure 2(a). For electric-field applied along 111 direction, Figures 4(a) and (b) illustrate changes in $G$ profile for polarization vectors within the $(010)$ and $(1\bar{1}0)$ planes, respectively. Figure 4(c) illustrates the polarization vectors for different possible domain variants, and their relative orientation to the applied electric field direction. In this case, application of electric-field creates a positive energy bias for certain domain variants. For example, the polarization directions of 001 and 100 within the (010) plane have lower free energy $G$ as compared to directions $00\bar{1}$ and $\bar{1}00$, as illustrated in Figure 4(a). Similarly, the polarization direction of 111 is most favored and has the lowest value of $G$ among all other directions within the $(1\bar{1}0)$ plane, as shown in Figure 4(b). The change in different domain variants, calculated based on eq.(2) and the relative free energy changes, are shown in Figures 4(d) and (e) for compositions, 52/48 and 50/50, respectively. In contrast to what is shown in Figure 3(d), Figure 4(d) shows that application of electric-field along the 111 direction favors a growth of the R1 variant of the rhombohedral phase, at the expense of other rhombohedral and tetragonal domain variants. However, for the 50/50 composition, the tetragonal domain variant T1 initially grows for lower values of electric-fields applied along 111 direction, followed by growth of the rhombohedral R1 domain variant at higher field magnitudes. The results shown in Figures 4(d,e) therefore illustrates that for grains that experience electric-field along the 111 direction, the rate of electric-field-induced transition from tetragonal to rhombohedral phase depends on the Zr/Ti ratio near the MPB. This is made clear from the plot shown in Figure 4(f), which indicates relative volume fractions of the rhombohedral and tetragonal phases as function of electric-field and for two different compositions.

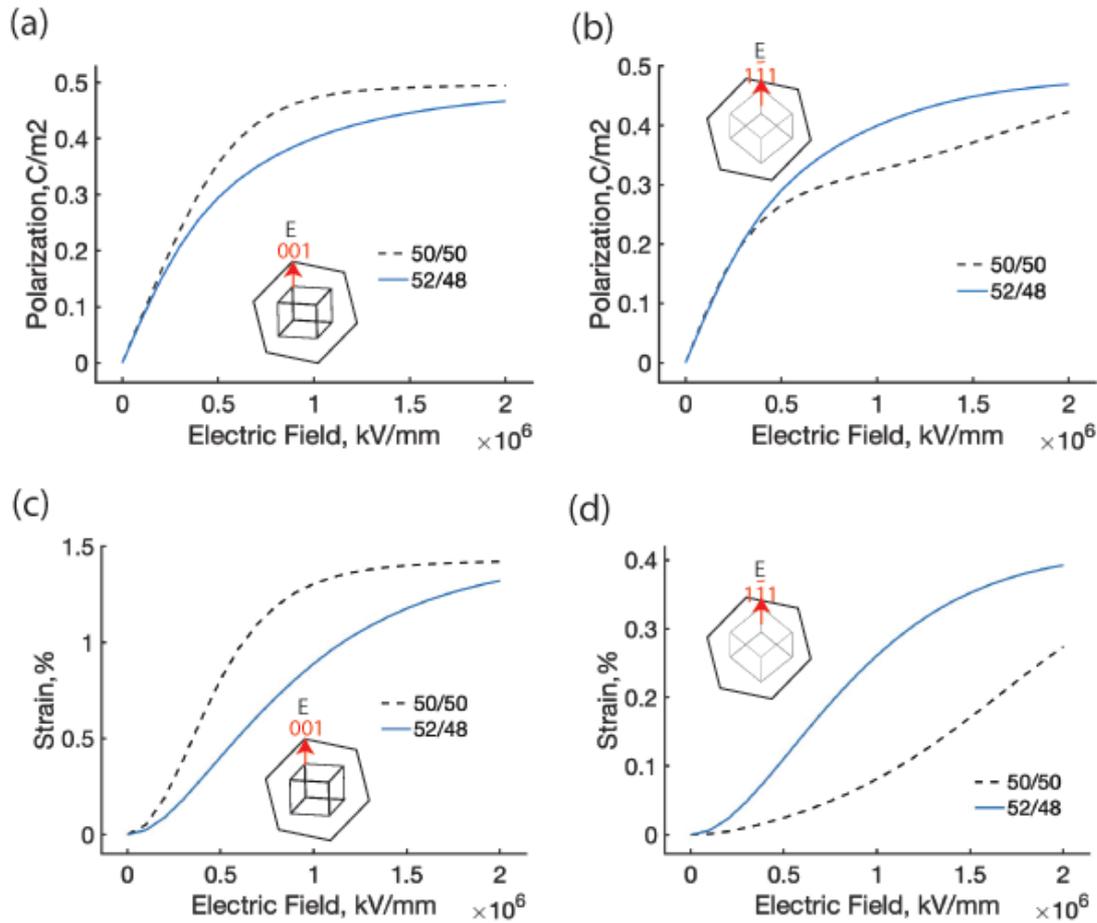

**Figure 5:** Polarization response of grains with two different orientations, (a) electric-field parallel to 001 direction, (b) electric-field parallel to 111 direction, plotted for two different values of Zr/Ti ratio. Strain response of grains with two different orientations, (c) electric-field parallel to 001 direction, (d) electric-field parallel to 111 direction, plotted for two different values of Zr/Ti ratio

The rate of different microscopic processes is expected to influence the electric-field-induced response of a material. For example, as shown in Figure 3, switching between tetragonal and rhombohedral ferroelectric phases plays a dominant role for the 52/48 composition, while switching between different domain variants within the tetragonal phase instead plays a more dominant role for the 50/50 composition. The consequence of this is illustrated in Figure 5(a), which shows the different polarization responses of grains with their 001 axes oriented parallel to the electric-field directions, for two different compositions of 52/48 and 50/50. Notably, grains of this orientation show a more gradual increase in polarization with increasing electric field for the 52/48 composition than that for the 50/50 composition. However, this situation changes for the grains with their 111 axis parallel to the electric-field direction. As shown in Figure 4, for this specific orientation, switching between tetragonal and rhombohedral phase variants is the predominant mechanism for both 52/48 and 50/50 compositions; however, this process progresses more gradually with increasing electric field in the latter case. The consequence of this is illustrated in Figure 5(b), which shows for grains with their 111 axis parallel to the electric-field direction, polarization increases more gradually with increasing electric-field for the 50/50 composition as compared to 52/48 composition.

Figure 5(c) shows the corresponding normal strain $\epsilon_{33}$ response of grains with their 001 axes oriented parallel to the electric-field direction, for two different compositions of 52/48 and 50/50. The method for calculating the total spontaneous strain matrix for a mixture of different domain variants is described in Appendix 1. The values plotted in Figure 5(c) indicate the strain for a certain grain relative to its state at zero electric -field. Figure 5(d) similarly shows the corresponding normal strain $\epsilon_{33}$ response of grains with their 111 axis oriented parallel to the electric-field direction, for two different compositions of 52/48 and 50/50. Note that for the above calculations, we considered polarization and strain changes as a result of domain or phase changes alone. For clarity, the induced polarization and strain responses due to linear dielectric and piezoelectric effects are not included, which are at least one order of magnitude lower than the corresponding effects due to domain/phase changes. Switching among the tetragonal domain variants produces a larger strain as compared to switching among the rhombohedral domain variants, due to a lower value of the spontaneous strain $S_0$ for the latter (see Appendix 1). Comparison of the results shown in Figures 5(c) and 5(d) indicates that the $\epsilon_{33}$ strain response measured along the 001 axis is larger for the 50/50 composition, while the $\epsilon_{33}$ strain response measured along the 111 axis is larger for the 52/48 composition.

3.2. Application of mechanical stress

In many different applications, ferroelectric ceramics are also subjected to mechanical stress, either solely or in conjunction with electric-fields.[42-44] For example, actuator components are likely to experience some pre-existing mechanical loads, even if unintentional. In addition, intergranular stresses are known to exist in polycrystalline ceramics.[45] Therefore, it will be interesting to know how much such stresses affect the volume fractions of the different domain/phase variants.

In order to identify the possible polar distortions under the application of an external stress, eq.(1) can be further expanded as follows:

$$\Delta G = \alpha_1[P_1^2 + P_2^2 + P_3^2] + \alpha_{11}[P_1^4 + P_2^4 + P_3^4] + \alpha_{12}[P_1^2 P_2^2 + P_2^2 P_3^2 + P_3^2 P_1^2] + \alpha_{111}[P_1^6 + P_2^6 + P_3^6] + \alpha_{112}[P_1^4(P_2^2+P_3^2) + P_2^4(P_3^2+P_1^2) + P_3^4(P_1^2+P_2^2)] + \alpha_{123} P_1^2 P_2^2 P_3^2 - {}^1\!/_2 S_{11}[X_1^2 + X_2^2 + X_3^2] - S_{12}[X_1 X_2 + X_2 X_3 + X_3 X_1] - {}^1\!/_2 S_{44}[X_4^2 + X_5^2 + X_6^2] - Q_{11}[X_1 P_1^2 + X_2 P_2^2 + X_3 P_3^2] - Q_{12}[X_1 (P_2^2+P_3^2) + X_2 (P_3^2+P_1^2) + X_3 (P_1^2+P_2^2)] - Q_{44}[X_4 P_2 P_3 + X_5 P_3 P_1 + X_6 P_1 P_2] \quad (3)$$

where $X_1,\ldots,X_6$ refer to the components of the stress tensor as $X_1 = \sigma_{11}$, $X_2 = \sigma_{22}$, $X_3 = \sigma_{33}$, $X_4 = \sigma_{23}$, $X_5 = \sigma_{13}$, $X_6 = \sigma_{12}$, $S_{ij}$ refer to the elastic compliances at constant polarization, $Q_{ij}$ refer to the electrostrictive coupling between the ferroelectric polarization and stress, and all other variables have the same meaning as described for eq.(1). The volume fractions of the different domain/phase variants are then calculated based on the value of $\Delta G$ under applied stress and applying equation (2) in the same manner as described in Section 3.1.

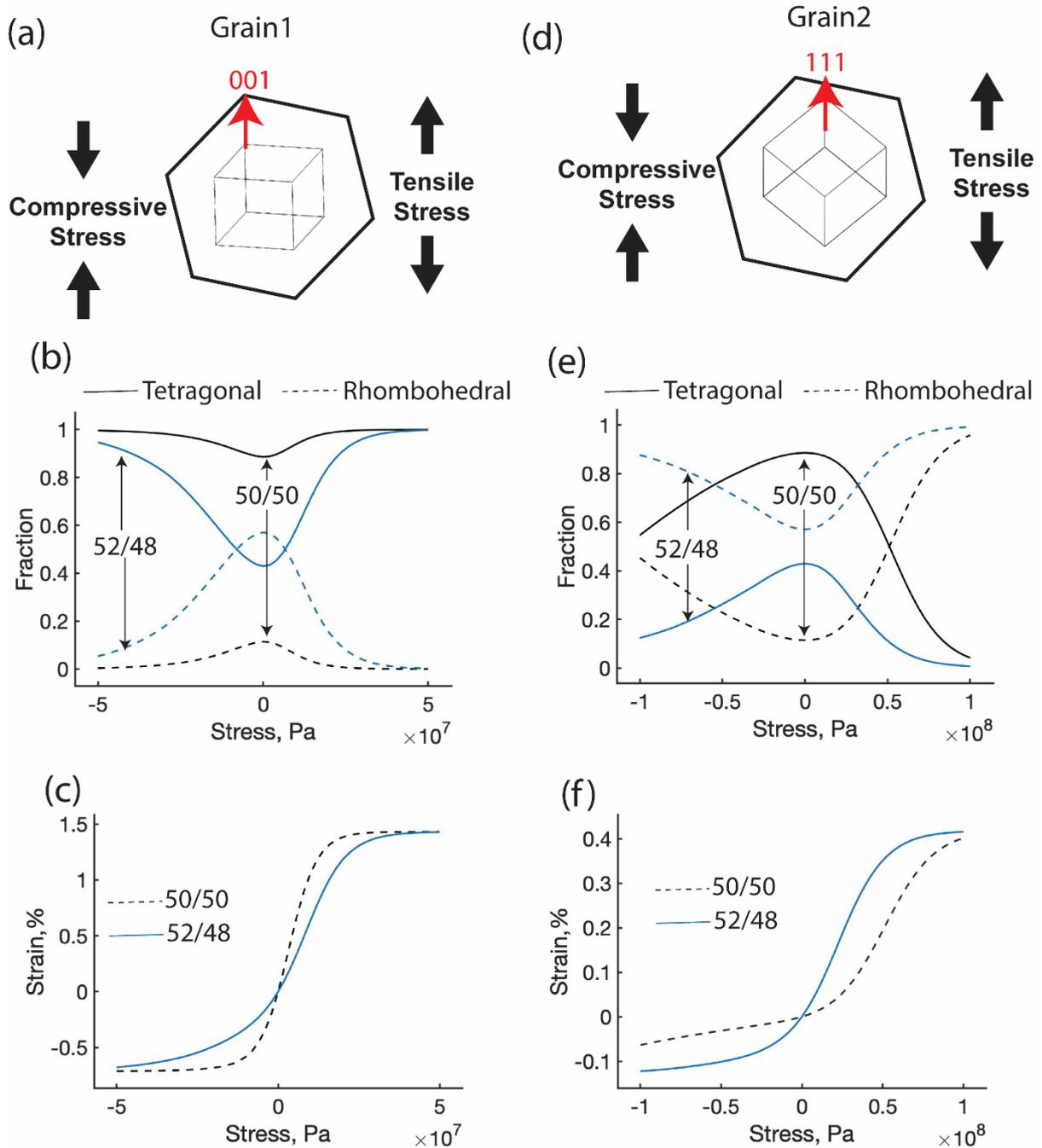

**Figure 6:** (a) Illustration of loading scenario for a grain with its 001 crystal axis parallel to that of the application of a tensile/compression stress. (b) Change in overall volume fractions of tetragonal and rhombohedral domains as function of applied stress along 001 and for two different values of Zr/Ti ratio. (c) Resultant strain as a result of domain switching and phase transitions for the loading scenario illustrated in (a). (d) Illustration of loading scenario for a grain with its 111 crystal axis parallel to that of the application of a tensile/compression stress. (e) Change in overall volume fractions of tetragonal and rhombohedral domains as function of applied stress and for two different values of Zr/Ti ratio. (f) Resultant strain as a result of domain switching and phase transitions for the loading scenario illustrated in (d).

We evaluated the effect of normal stress, $X_3$, applied parallel to either 001 or 111 crystal axes, and for two different compositions 52/48 and 50/50, as illustrated in Figure 6(a,d). Figure 6(b) shows changes in volume fractions of the tetragonal and rhombohedral domain variants under tensile and compressive stresses. It shows that application of either tensile or compressive stress favors an increase in tetragonal domain variants at the expense of rhombohedral domain variants. Similarly, Figure 6(e) shows that application of normal tensile or compressive stress, $X_3$, along 111 crystal axis favors rhombohedral domain variants at the expense of tetragonal domain variants. In both cases, there exists asymmetry whereby phase transition is stronger under tension than the same under compression. The resulting strains arising from ferroelastic domain-switching/phase transition under tensile/compressive stress applied along 001 and 111 crystal axes are shown in Figures 6(c) and (f), respectively. Note that linear elastic strain resulting from application of stress is not included in this calculation for clarity, but they can be superimposed to the corresponding ferroelastic and phase transition strains. For stress applied parallel to 001 crystal axis, the ferroelastic strains under both tensile and compressive stress are larger for the composition 50/50 than the same for composition 52/48. In comparison, for stress applied parallel to 111 crystal axis, the strains under both tensile and compressive stress are larger for the composition 52/48 than the same for composition 50/50. In addition, the saturation strain is also larger for tension as compared to that for compression. This reflects the fact that for stress applied along 001 and for a positive value of spontaneous strain, a maximum of 2/3 of the tetragonal non-180° domain variants can be reoriented under tensile stress, while a maximum of 1/3 of the tetragonal non-180° domain variants can be reoriented under compressive stress, relative to the zero stress state. Similarly, for stress applied along 111 and for a positive value of spontaneous strain, a maximum of 3/4 of the rhombohedral non-180° domain variants can be reoriented under tensile stress, while a maximum of 1/4 of the rhombohedral non-180° domain variants can be reoriented under compressive stress, relative to the zero stress state.

3.3. Simultaneous application of mechanical stress and electric field

In reality, ferroelectric devices often experience a combination of mechanical loads and electric-fields. [42-44] Even when only electric-field is applied to a ferroelectric ceramic component, differently oriented grains can experience intergranular stresses.[45] It is therefore of interest to examine how domain switching and phase transition proceed under such combined electromechanical loading scenario. We examine here a particular case of application of incremental electric-field under constant compressive stress for the MPB composition 52/48, and for two different crystal axes, as depicted in Figure 7(a,e). For evaluation of the Landau free energy under combined electromechanical loading, we combine equations (2) and (3). Thereafter, the procedure followed to evaluate the equilibrium volume fractions of the different domain variants is the same as described above.

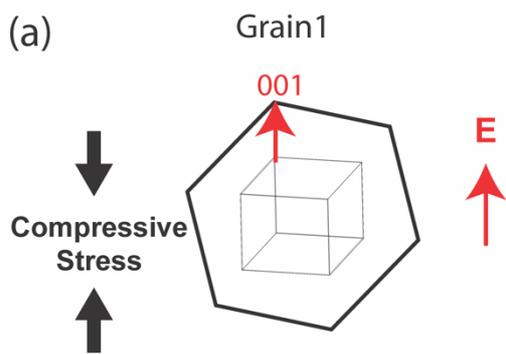
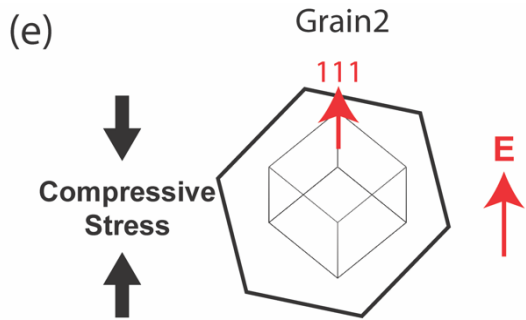
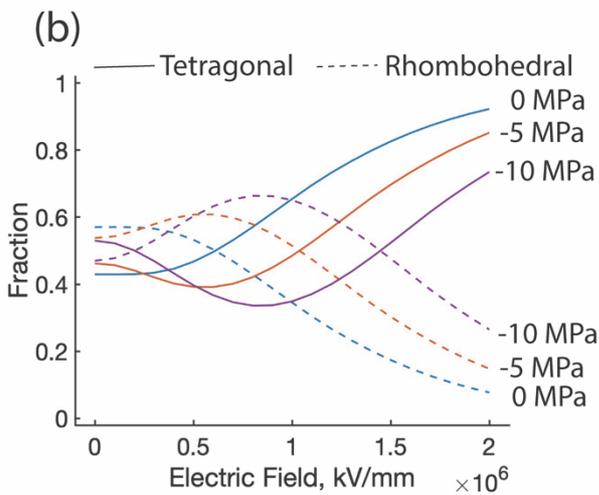
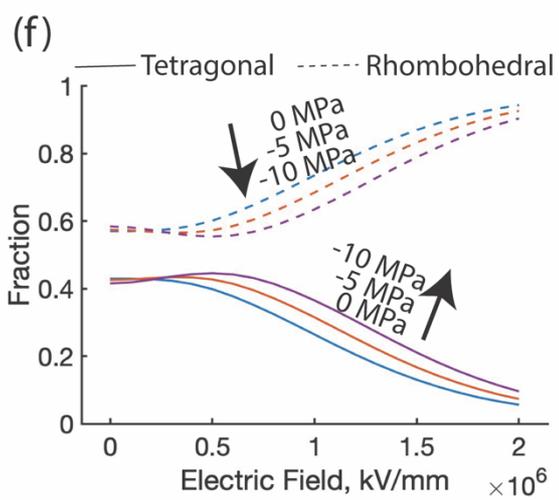
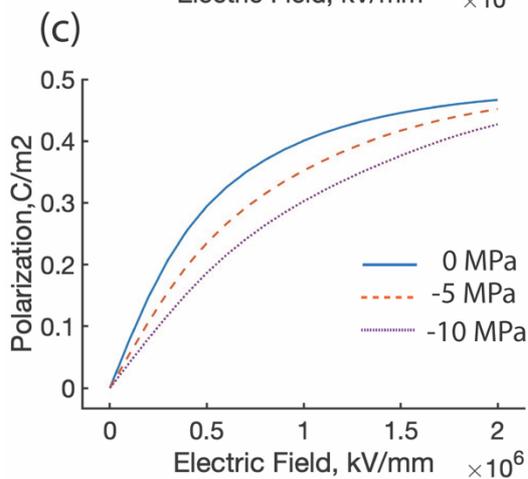
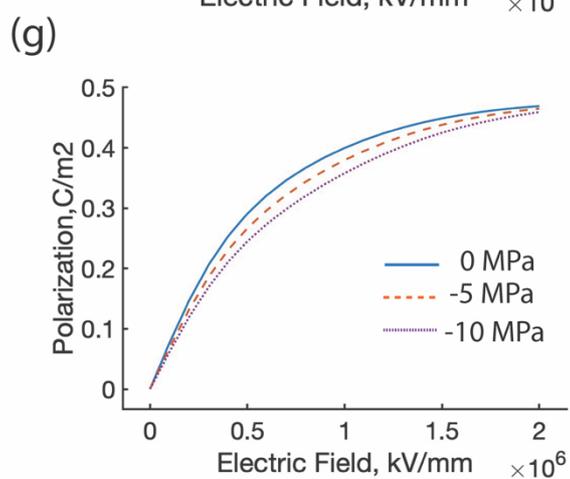
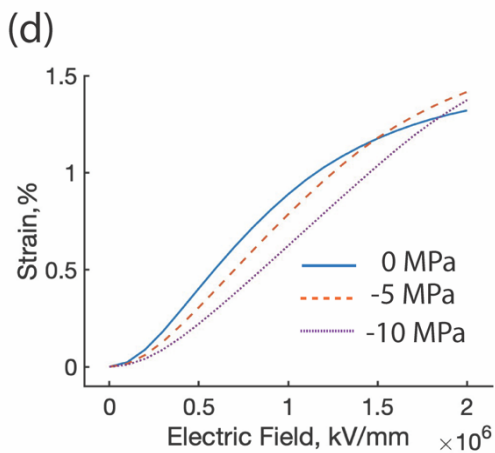
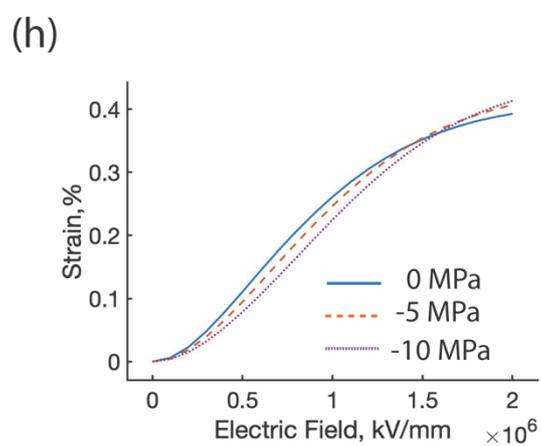

**Figure 7:** (a) Illustration of loading scenario for a grain with its 001 crystal axis parallel to that of the application of simultaneous electric field and compressive stress. (b) Change in overall volume fractions of tetragonal and rhombohedral domains as function of incremental electric field and constant compressive stress applied along 001 crystal axis.( c,d) Resultant polarization and strain as a result of domain switching and phase transitions for the loading scenario illustrated in (a). (e) Illustration of loading scenario for a grain with its 111 crystal axis parallel to that of the application of simultaneous electric field and compressive stress. (f) Change in overall volume fractions of tetragonal and rhombohedral domains as function of incremental electric field and constant compressive stress applied along 111 crystal axis. (g,h) Resultant polarization and strain as a result of domain switching and phase transitions for the loading scenario illustrated in (e).

For simultaneous application of mechanical stress and electric field along the 001 crystal axes, the change in volume fraction of tetragonal and rhombohedral phases are shown in Figure 7(b). Under increasing magnitudes of compressive stress, the volume fraction of the tetragonal phase decreases, while that of the rhombohedral phase increases. Interestingly, the volume fraction of the tetragonal (rhombohedral) phase initially decreases (increases) with the application of smaller electric fields, before eventually increasing (decreasing) under higher electric fields. The resultant polarization and strain responses are shown in Figures 7(c) and (d), respectively. The electric-field-induced polarization is lower for increasing magnitudes of compressive stress under intermediate electric-field values, although saturation polarization is approached under higher electric-fields. Similar behavior is observed for the induced strain due to combined domain-switching and phase transition mechanisms, as shown in Figure 7(d). Note that the apparent strain is slightly larger at higher electric-fields under compressive stress, which is due to the fact that the overall strain is computed relative to the initial state and the material is initially compressed under zero electric-field.

Figures 7(f,g,h) show changes in phase volume fractions, polarization and strain responses under simultaneous application of compressive stress and electric

field parallel to the 111 crystal axis. In this case, as shown in Figure 7(f), the volume fraction of the tetragonal phase increases, while that of the rhombohedral phase decreases under increasing magnitudes of compressive stress. Additionally, both electric-field-induced polarization and strain values are lower under intermediate electric-field values and for increasing magnitudes of compressive stress, as shown in Figures 7(g,h); however, saturation is approached in both cases under higher electric-fields. Note that the effect of a compressive stress on electric-field-induced polarization/strain is relatively weaker for the situation depicted in Figure 7(b), as compared to the same for what is shown in Figure 7(a). This is caused by how the differences between the energy minima of the different domain variants are relatively affected under different electromechanical loading scenarios.

3.4. Generalization to all orientation states

At this point, we can generalize the methodology to compute the extent of phase transition for any relative misorientation between the crystal axes and the axes of applied electric field and/or stress. A convenient way to represent the results is through the use of inverse pole figures, as shown in Figure 8. The stereographic projections are made with respect to the <001> cubic axes as indicated in the Figure 8. The various points on the stereographic projection indicate the directions of the applied electric field or the axis of applied uniaxial stress with respect to the crystallographic pseudocubic 001 crystal axes.

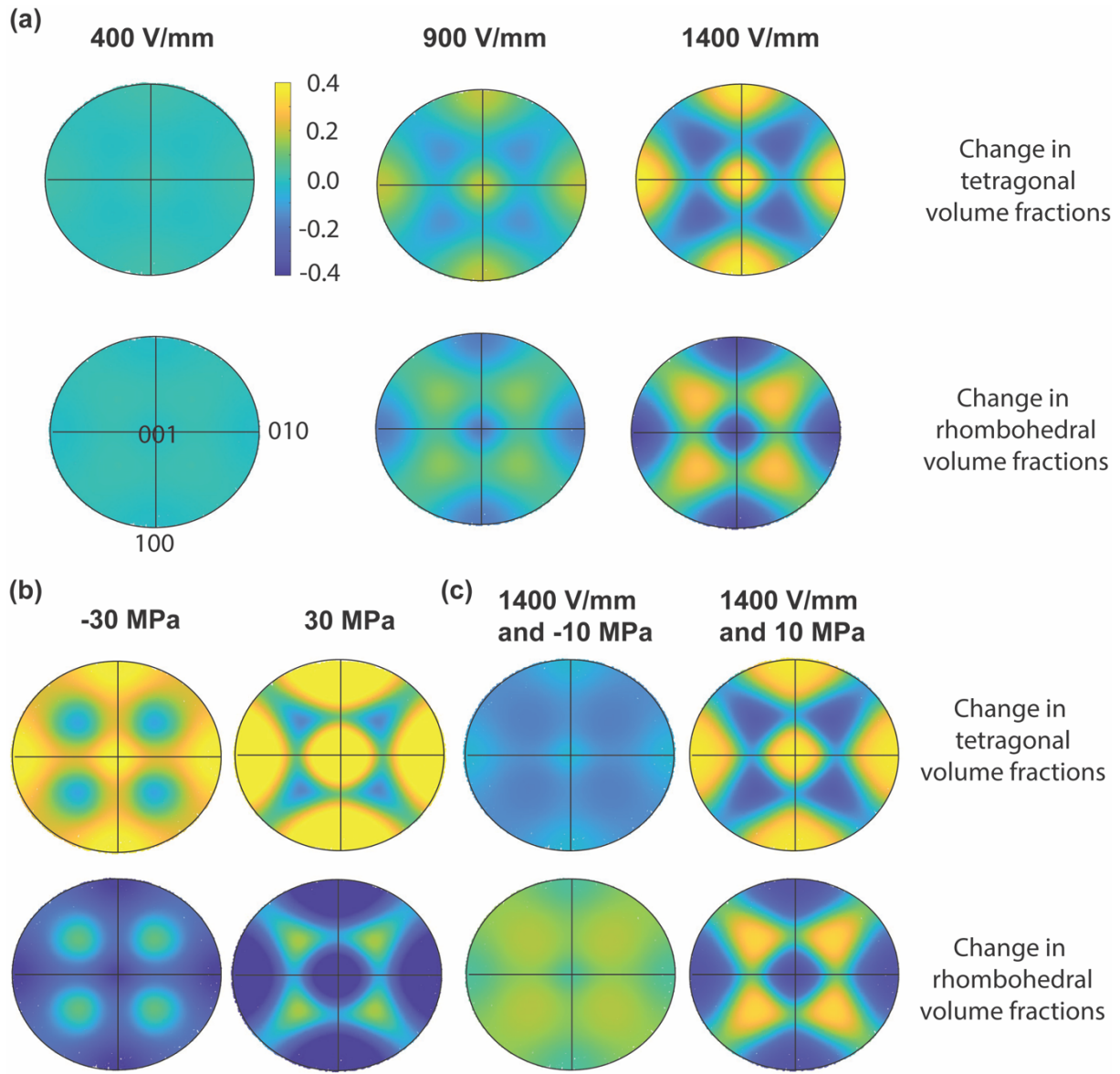

**Figure 8:** Inverse pole figures showing degree of phase change for applied electric-field and/or uniaxial stress parallel to different crystallographic directions. (a) Change in tetragonal and rhombohedral phase fractions for different electric field directions and magnitudes. (b) Same as in (a), for applied uniaxial tensile and compressive stresses. (c) Same as in (a), for simultaneous application of co-aligned electric field and uniaxial stress.

The top two panels in Figure 8(a) indicate changes in the tetragonal and rhombohedral volume fractions for different directions of applied electric fields and for different field magnitudes. The changes in phase volume fractions become more prominent for applied electric fields higher than 1 kV/mm. This is consistent with the general experimental observations, such as [46]. The plots indicate the different regions within the orientation space, which favors an increase/decrease in either the tetragonal or rhombohedral phase volume fractions. The tetragonal phase fraction increases for electric field directions closer to <001> crystal axes, while it decreases for electric field directions closer to <111> crystal axes. Conversely, the rhombohedral phase fraction increases for electric field directions closer to <111> crystal axes, while it decreases for electric field directions closer to <001> crystal axes.

Figure 8(b) indicates the extent of changes in relative phase fractions for applied tensile/compressive stresses and for combined application of electric-field and stress. In the case of applied compressive stress, the separation between the regions with preference for either tetragonal or rhombohedral phases is more diffused as compared to what is observed for applied electric field alone or for applied tensile stress alone. In the case that electric field and stress are applied simultaneously, a convolution of the effects observed for each of these cases can be expected, as depicted in Figure 8(c). For simultaneous application of positive electric field and compressive stress, the boundaries between the regions with different phase preferences are very diffused and rhombohedral phase has a higher preference in general. In contrast, for simultaneous applied positive electric field and tensile stress, the different regions in the orientation space with clear preference for tetragonal and rhombohedral phases are more clearly defined.

## 4. Application to polycrystalline aggregates

Once the behavior of single crystallites for all different orientations (with respect to macroscopic electric-field and/or stress) are determined, we can proceed to estimate the functional response of a polycrystalline aggregate by averaging over all orientation states. In a simple approach, the electric-field and stress can be assumed to be uniformly applied to each single crystallite in a polycrystalline ensemble. In this case, one can obtain the response of the polycrystalline aggregate by simply averaging over all the orientation states, such as adopted in refs. [47,48]. In a further refinement, one can take into account the grain-to-grain interactions and re-optimize the level of localized electric-field and stress that are internally experienced by each single crystallite; such models have been used for example in refs. [27,33,49]. Here, we use the former approach, that is, we consider the electric-field and stress to be uniform for all single-crystal grains, as a first-approximation to characterize the progressive evolution of domain and phase volume fractions in a material. The effect of field and stress localization will be provided in a subsequent article.

*(a) Domain and phase switching*

First, we simulate the microstructure of a polycrystalline ceramic with 20,000 grains with a random texture. This is done by assigning random values to the three Euler angles $\phi_1$, $\psi$ and $\phi_2$, as described in Appendix 2. The computed results are presented by grouping the different orientations in terms of their azimuhtal angles $\psi$ into bins of 10° intervals. This is done to correlate the modelling results with the results of *in situ* X-ray diffraction measurements, which typically follow this convention, such as refs. [45,50-52]

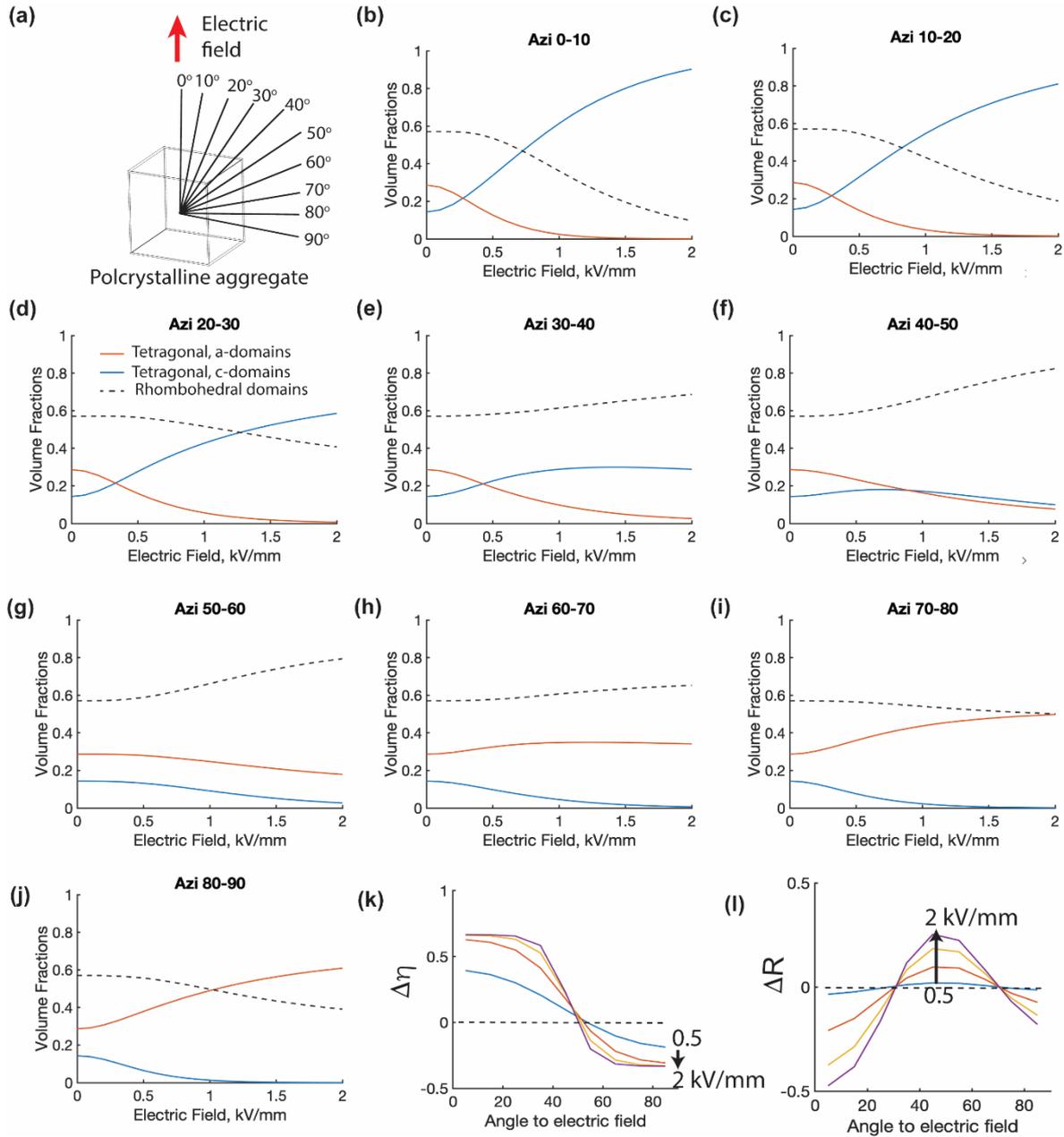

**Figure 9:** (a) Definition of the different azimuthal angle bins with respect to the electric field direction. Note that the material response is considered to be transversely isotropic with respect to the electric field direction. (b-j) Electric field dependent changes in volume fractions of *a*- and *c*-domains of the tetragonal phase and the rhombohedral phase, for polycrystalline PZT with Zr/Ti=52/48. (k) $\Delta\eta$, which signifies extent of 90° domain switching in the tetragonal phase, is presented for various azimuthal angular bins. (l) $\Delta R$, which signifies extent of tetragonal-to-rhombohedral phase transition, is presented for various azimuthal angular bins.

Figure 9 shows the changes in volume fractions of the *c*- and *a*-domains of the tetragonal phase, as well as the rhombohedral phase volume fraction, for various azimuthal ($\psi$) angle bins, for composition with Zr/Ti ratio of 52/48. The various azimuthal angular bins with respect to electric field direction are indicated in Figure 9(a). Within the azimuthal angular range of 0°-40°, the volume fraction of the *c*-domains increases while that of the *a*-domains decrease; the opposite is observed for the angular range of 70°-90°. For the azimuthal angular range of 30°-70°, the relative volume fraction of the *a*- and *c*-domains exhibit relatively smaller changes. Additionally, for the angular range of 30°-70°, an increase in the rhombohedral phase volume fraction with applied electric field can be observed, especially for electric fields higher than 1 kV/mm. In order to numerically represent the microstructural changes, we further compute the following two quantities:

$$\eta = \frac{V_{002}}{V_{002}+V_{200}} \quad (4)$$

$$R = \frac{V_R}{V_T+V_R} \quad (5)$$

where $V_{002}$ and $V_{200}$ represent the *c*- and *a*-domain volume fractions of the tetragonal phase, respectively; $V_T$ and $V_R$ represent the volume fractions of the tetragonal and rhombohedral phases, respectively. $\Delta\eta$ and $\Delta R$ represent changes in the values of these variables, with respect to their original values at zero applied electric fields. In effect, $\Delta\eta$ represents degree of 90° domain switching and $\Delta R$ represents tetragonal↔rhombohedral phase switching under the application of electric-field and/or stress. Figure 9(k) shows $\Delta\eta$ as a function of $\psi$ for various magnitudes of applied electric fields. The dotted line indicates the baseline of $\Delta\eta$ = 0 for zero applied electric field. The computed angular distribution for $\Delta\eta$ is consistent with that of experimental measurements reported for PZT ceramics near MPB composition [45,50,51].

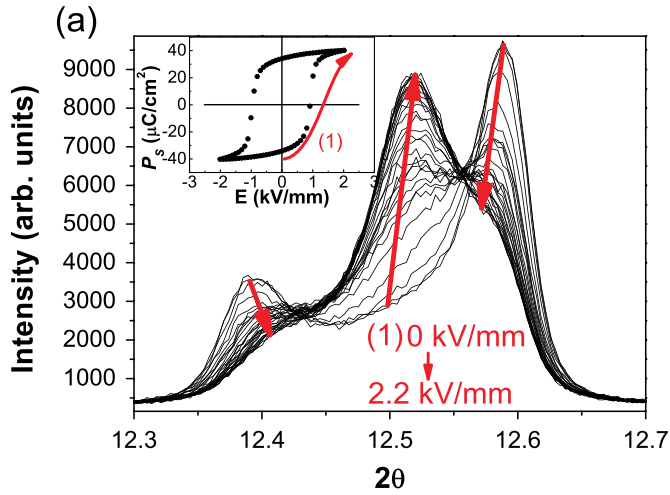

(a)

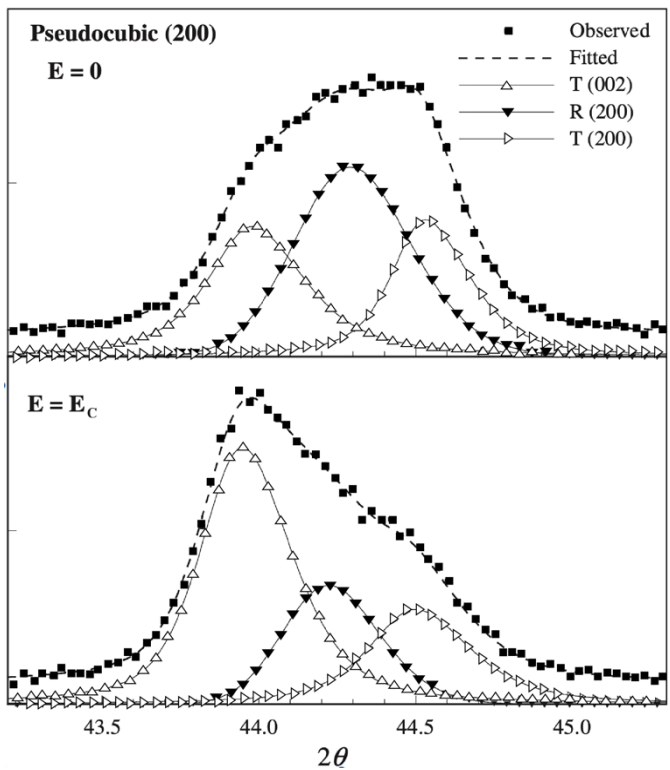

(b)

**Figure 10**: (a) Change in 002 diffraction peak profile from a PZT ceramic of MPB composition under the application of electric field, **E,** which is rotated 45° from 002 diffraction wavevector; reproduced with permission from [46] (b) Change in 002 diffraction peak profile from a PZT ceramic of MPB composition under the application of electric field, **E**, which is parallel to 002 diffraction wavevector; reproduced with permission from [54]

Figure 9(I) similarly show angular distribution of $\Delta R$ for various applied electric field magnitudes. The results provide a phenomenological understanding of experimental reports of electric-field-induced phase transformations for materials near MPB. Hinterstein *et al.*[46] reported change in X-ray diffraction patterns for PZT ceramics, which indicated electric-field-induced transformation from tetragonal to monoclinic phase. For example, Figure 10(a) shows changes in {002} diffraction peak profile for electric-field applied along direction that is rotated 45° from the 002 diffraction wavevector.[46] The appearance of a third peak in-between the two principal 002/200 reflections from the tetragonal domain variants was explained as a result of an induced monoclinic phase. Similar results were also reported for the solid-solution system of $PbTiO_3$-$BiScO_3$,[53] which exhibits a an MPB similar to PZT. These experimental results validate our modelling, whereby one can observe a peak in positive values of $\Delta R$ within the angular range of $\psi = 40° - 50°$, as shown in Figure 9(l). Note that, in our modelling, we have considered the polarization states near <111> directions as "rhombohedral"; however, the flatness of the energy profile means that small deviations from ideal rhombohedral states with <111> polarization directions can occur, which can be interpreted as "monoclinic". For $\psi = 0° - 30°$, our modelling indicates large negative values of $\Delta R$, that is, a rhombohedral-to-tetragonal phase transition for grain of these orientations. This is also validated by results from *in situ* XRD experiments, such as reported by ref.[54] and shown in Figure 10(b). As shown in Figure 10(b), for a PZT ceramic of MPB composition, the rhombohedral domain variants have a larger contribution to the overall 002 diffraction peak intensity at zero electric-field, as compared to that under electric field **E** that is applied approximately parallel to 001 crystallographic poles.

    The domain and phase switching behavior in polycrystalline PZT with Zr/Ti ratio of 50/50 is shown in Figure 11. For Zr/Ti=50/50, the volume fraction of the rhombohedral phase is much lower as compared to that observed for Zr/Ti = 52/48. However, the angular distribution of $\Delta\eta$ observed for Zr/Ti=50/50 is similar to what is observed for Zr/Ti = 52/48. This is not surprising since the *change* in free energies of

the polarization states for the *a*- and *c*-domains remains unchanged with composition, although the absolute values of their volume fractions are changed. In addition, we can note from Figure 11(l) that the extent of rhombohedral-to-tetragonal phase transition is reduced as compared to Zr/Ti = 52/48, while the angular range for tetragonal-to-rhombohedral phase transition is broadened.

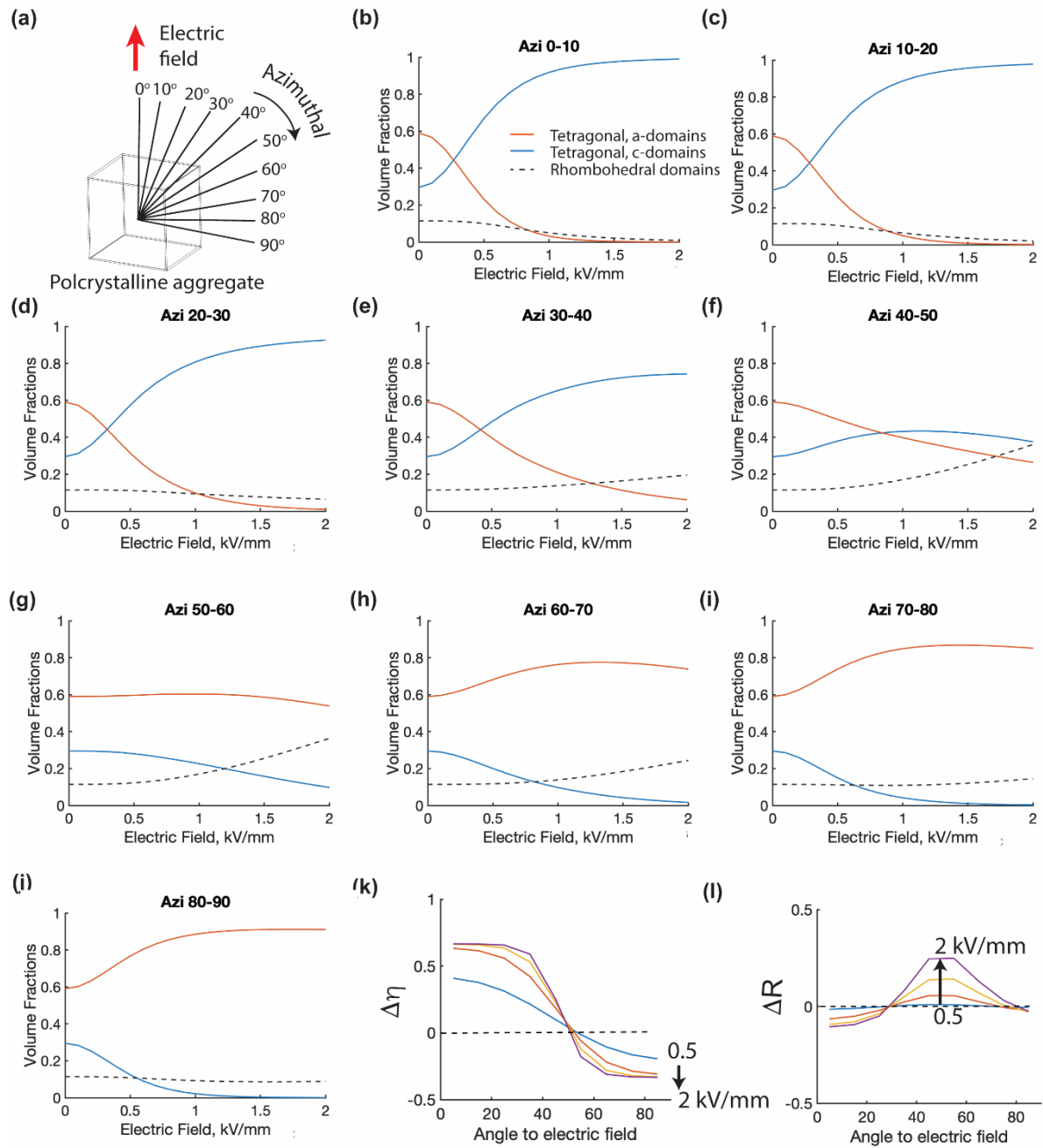

**Figure 11**: Same as in Figure 9, but for Zr/Ti ratio equal to 50/50.

The domain and phase switching behavior can be also subject to change under simultaneous application of electric field and stress. An example of this is illustrated in Figure 12, which shows electric field induced evolution of various domain volume fractions for different azimuthal angular range under simultaneous application of electric field and a constant compressive stress of -10 MPa. Interestingly, for $\psi = 0° - 30°$ and $\psi = 70° - 90°$, one can observe an initial increase in the rhombohedral phase fraction for electric field magnitudes lower than 1 kV/mm, followed by subsequent decrease of the same under higher electric fields. The angular distribution of $\Delta\eta$ and $\Delta R$ are shown in Figure 12(k) and (l), respectively. It shows that $\Delta\eta$ or domain switching is initially suppressed for field magnitudes below 1 kV/mm, but then accelerates under higher electric fields. Also, for angles closer to the electric field direction, there is an increase in the rhombohedral phase fraction for electric fields lower than 1 kV/mm, which is subsequently reversed for higher electric field magnitudes.

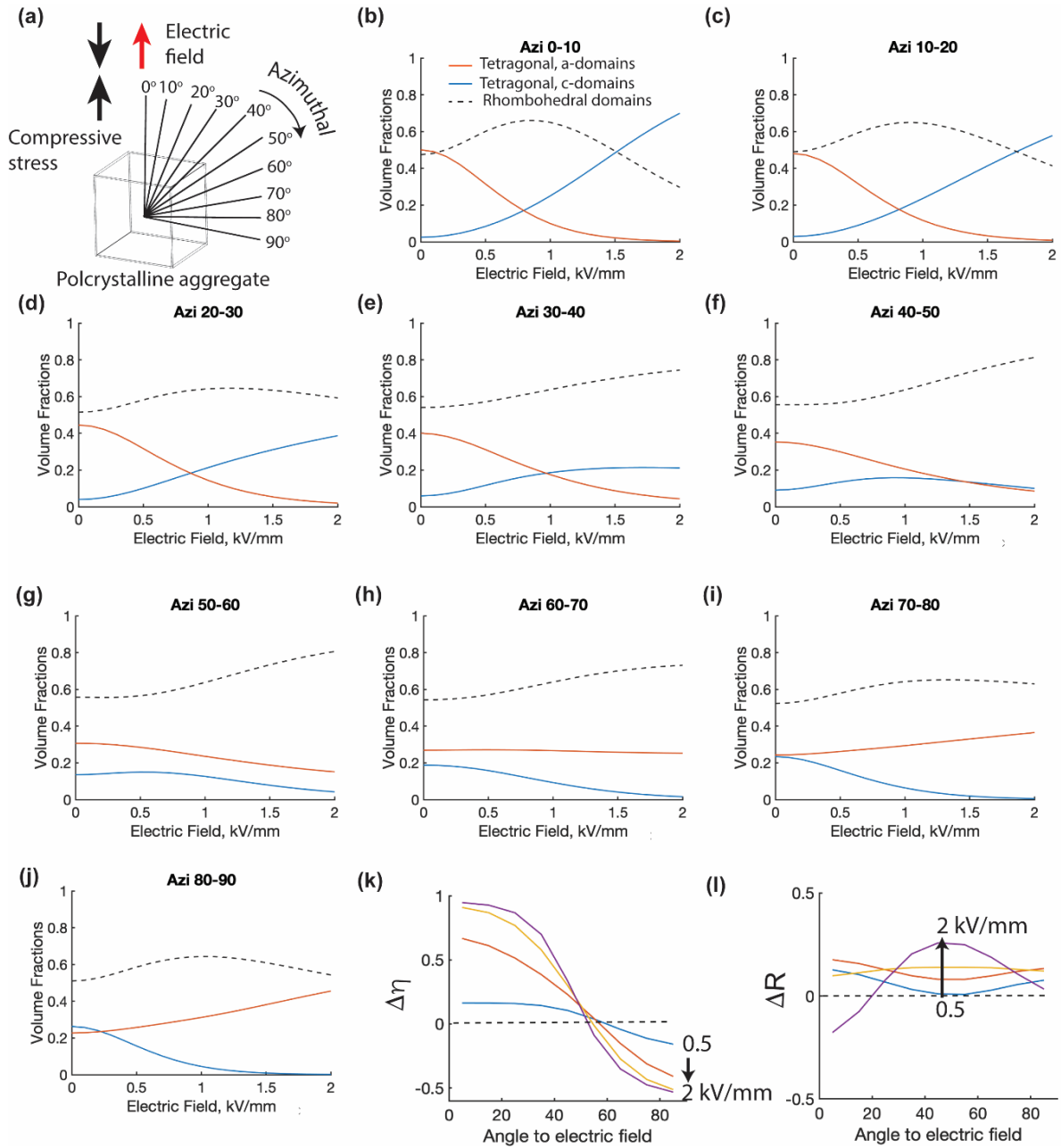

**Figure 12:** Same as in Figure 9, but for simultaneous application of electric field and a constant compressive stress of -10 MPa.

*(b) Contribution to macroscopic response*

The modelling results presented above, correlates with earlier experimental reports, and indicate that grains with different orientations within a polycrystalline aggregate can exhibit different domain/phase switching behavior depending on their specific orientations with respect to macroscopic field/stress direction. Depending on specific grain orientations, the dominant microscopic mechanism can be either 90° domain switching, rhombohedral-to-tetragonal or tetragonal-to-rhombohedral phase transition. Therefore, a pertinent question is which of these mechanisms is the most significant contributor to the macroscopic functional response of a polycrystalline ceramic. In order to answer this, we next calculate the relative contributions of the grains with different orientations to the macroscopic material response, based on the results of $\Delta\eta$ and $\Delta R$, such as presented in Figures 9, 11 and 12. Once the values of $\Delta\eta$ and $\Delta R$ are calculated for a specific grain orientation, the corresponding polarization and strain contributions from domain/phase switching are obtained through a tensorial transformation as explained in the Appendix 2.

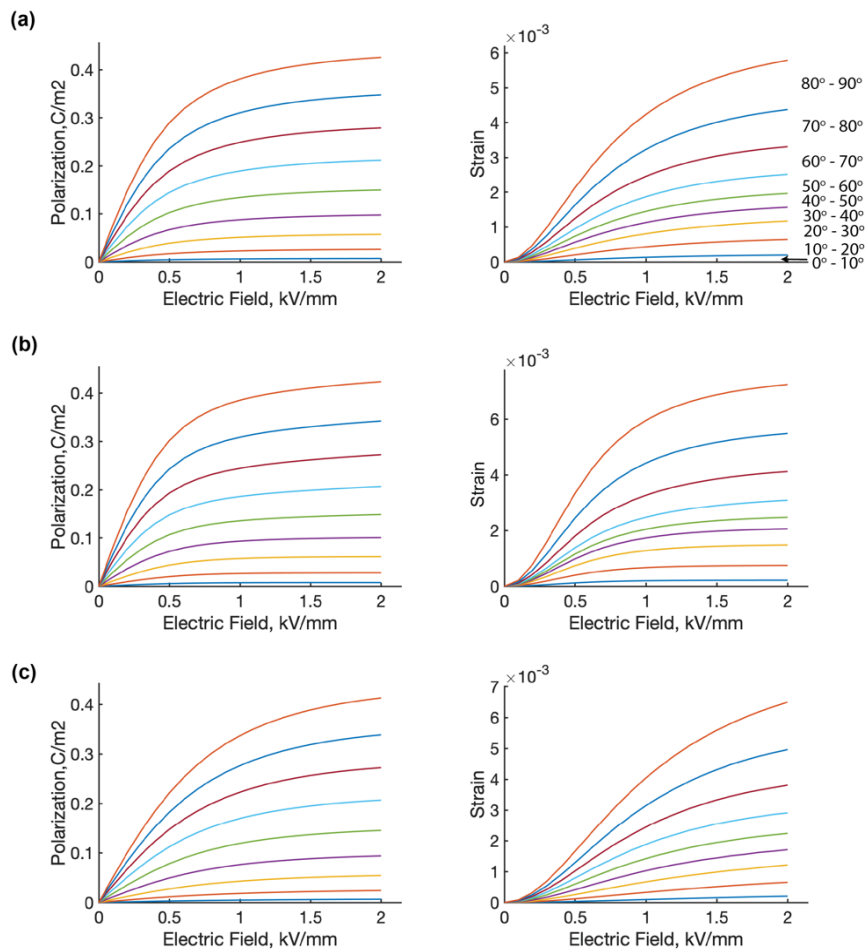

**Figure 13:** Contributions to macroscopic electric-field induced polarization and strain due to phase and domain switching phenomena from grains with different azimuthal orientations with respect to electric field or stress direction, as defined in Figure 9. (a)

for Zr/Ti = 52/48 under applied electric fields, (b) for Zr/Ti = 50/50 under applied electric fields, and (c) for Zr/Ti = 52/48 under applied electric fields and constant compressive stress of -10 MPa.

Figure 13 shows the contributions to the macroscopic polarization and strain from single crystal grains of different orientations, which are defined with respect to the range of azimuthal $\psi$ angles, as defined in Appendix 1. It is convenient to define the relative contributions from grain of different orientations based on their relative azimuthal orientation angle $\psi$, since a polycrystalline ceramic can be assumed to be transversely isotropic with respect to an applied electric field or uniaxial stress direction. The azimuthal sectors represented in Figure 9 also correlate well with the convention typically used for *in situ* X-ray diffraction measurements, such as in [51]. Figure 13(a,b) compares the polarization and strain contributions from grains of different azimuthal orientations for the two different compositions of Zr/Ti = 52/48 and Zr/Ti = 50/50. Figure 13(c) compares the contributions to polarization and strain from grains of different azimuthal orientations for the composition 52/48 under applied electric fields and a constant applied stress of -10 MPa.

Furthermore, we can separate the relative contributions from grains with different orientations into three separate groups: (a) group I for $\psi = 0° - 30°$ (~13% of grains for random texture) that exhibits positive(negative) values for $\Delta\eta$ ($\Delta R$), (b) group II for $\psi = 30° - 70°$ (~52% of grains for random texture) that exhibits positive values for $\Delta R$ and relatively smaller $\Delta\eta$, and (c) group III for $\psi = 70° - 90°$ (~35% of grains for random texture) that exhibits negative values for both $\Delta\eta$ and $\Delta R$. Groups I and III represent grain orientations in which 90° domain switching and rhombohedral-to-tetragonal phase transition are the dominant microscopic mechanisms. Group II represents grain orientations in which tetragonal-to-rhombohedral is the dominant microscopic mechanism. In all cases, the relative contributions to macroscopic polarization from grains of Groups I, II and III roughly scale with their relative volume fractions. However, it is not the case for their strain contributions.

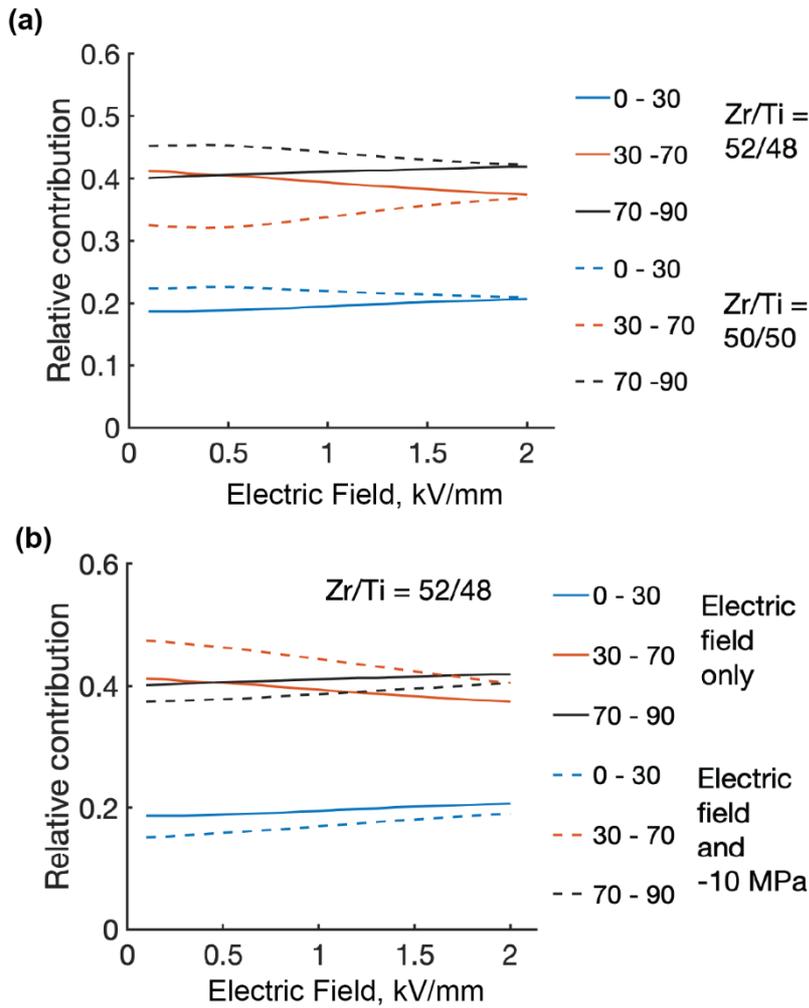

**Figure 14:** Comparison of relative contributions to macroscopic strain from grains with different azimuthal orientations range for (a) Zr/Ti ratio of 52/48 and 50/50, (b) Zr/Ti ratio of 52/48 under combined applied electric fields and constant compressive stress.

The relative contributions to strain from grains of Groups I,II,III are depicted as function of electric field in Figure 14. The contributions to macroscopic strain from grains of Group II for small electric -fields is ~0.41 and ~0.32 for compositions of 52/48 and 50/50, respectively, however, both these values approach ~0.37 for an electric field of 2 kV/mm. This provides an estimate of the relative contributions to overall strain from grains in which tetragonal-to-rhombohedral phase transition constitutes a major electric-field-induced microscopic phenomenon. The remaining contribution comes from grains in which 90° domain switching and rhombohedral-to-tetragonal phase transition are the dominant microscopic mechanisms, that is Group I and Group II. It is notable that the relative contribution from Group II grains increases under a compressive stress as compared to zero stress condition, however, it also decreases with increasing electric fields. Overall, for PZT ceramics of compositions slightly on the tetragonal side of the MPB, we can estimate that: (a) grains with net tetragonal-to-rhombohedral phase transition (~52% of grains for random texture) contributes roughly 30-40% of the overall strain, and (b) grains with net rhombohedral-to-tetragonal phase transition and large 90° domain switching (~48% of grains for random texture) contributes roughly 60-70% of the overall strain.

## 4. Concluding remarks

In summary, we presented here a generic model based on a combination of LGD phenomenological theory and Boltzmann statistics, which can be used to predict domain switching and/or phase transition in complex ferroelectrics with multiple coexisting polar phases. The methodology allows for the examination of material behavior under different loading conditions, such as electric-field and/or stress, as well as taking into account relative grain orientations. Additionally, it is possible to incorporate more complex scenarios such as temperature variations, as long as the temperature dependence of the Landau coefficients are known.

The results presented here are noteworthy in underlining the significance of specific grain orientations while discussing the relative importance of either domain switching or phase transition phenomenon towards the functional response of a ferroelectric material. The results indicate that generalizations relating enhanced properties in ferroelectrics of certain composition to solely domain switching or phase transition need to be properly examined. In order to interpret the macroscopic response of a polycrystalline material, it is necessary to consider grains of all possible orientations and weighing their relative contributions to the macroscopic behavior. The methodology for the same is provided here in the framework of a polycrystalline aggregate with random texture, which allows for the estimation of relative contributions from families of grains that undergo different degrees of phase transition and domain switching phenomena. An important result from our model is that, for PZT composition near MPB, the grain families that exhibit larger 90° domain switching and a net rhombohedral-to-tetragonal phase switching have a larger contribution to macroscopic electric-field induced strain, as compared to grains that exhibit net tetragonal-to-rhombohedral phase switching.

**Acknowledgements**

Support from CentraleSupelec and University Paris-Saclay in the framework of the d'Alembert fellowship program is gratefully acknowledged

# Appendix

Appendix 1: Calculation of domain-switching strains

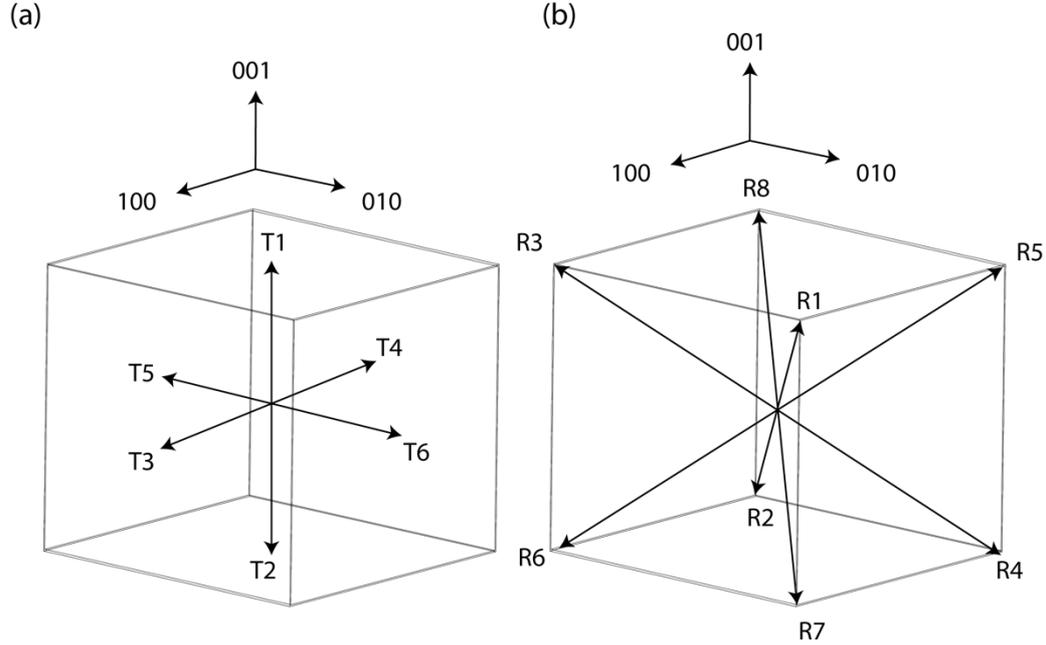

Figure A.1: Definition of the polarization directions for (a) tetragonal domains, (b) rhombohedral domains, with respect to pseudocubic crystallographic coordinates

In a tetragonal crystal, the six possible polarization directions are along the <001> pseudo-cubic crystal axes, which are indicated in Figure 9(a). The spontaneous strains corresponding to each of these six possible polarization directions are as follows [49]:

$$\varepsilon^{T1} = \epsilon^{T2} = \frac{S_0}{3}\begin{bmatrix} -1 & 0 & 0 \\ 0 & -1 & 0 \\ 0 & 0 & 2 \end{bmatrix}$$

$$\varepsilon^{T3} = \epsilon^{T4} = \frac{S_0}{3}\begin{bmatrix} 2 & 0 & 0 \\ 0 & -1 & 0 \\ 0 & 0 & -1 \end{bmatrix}$$

$$\varepsilon^{T5} = \epsilon^{T6} = \frac{S_0}{3}\begin{bmatrix} -1 & 0 & 0 \\ 0 & 2 & 0 \\ 0 & 0 & -1 \end{bmatrix} \qquad (1)$$

where $S_0 = \left(d_{001}/d_{100} - 1\right)$, $d_{001}$ and $d_{100}$ corresponds to 001 and 100 plane spacings, respectively. For tetragonal domains, we used $S_0 = 2.15\%$

In a rhombohedral crystal, the eight possible polarization directions are along the <111> pseudo-cubic crystal axes, which are indicated in Figure 9(b). The spontaneous strain corresponding to each of these eight possible polarization directions are as follows [50]:

$$\varepsilon^{R1} = \epsilon^{R2} = \frac{S_0}{3}\begin{bmatrix} 0 & 1 & 1 \\ 1 & 0 & 1 \\ 1 & 1 & 0 \end{bmatrix}$$

$$\varepsilon^{R3} = \epsilon^{R4} = \frac{S_0}{3}\begin{bmatrix} 0 & -1 & 1 \\ -1 & 0 & -1 \\ 1 & -1 & 0 \end{bmatrix}$$

$$\varepsilon^{R5} = \epsilon^{R6} = \frac{S_0}{3}\begin{bmatrix} 0 & -1 & 1 \\ -1 & 0 & 1 \\ 1 & 1 & 0 \end{bmatrix}$$

$$\varepsilon^{R7} = \epsilon^{R8} = \frac{S_0}{3}\begin{bmatrix} 0 & 1 & -1 \\ 1 & 0 & -1 \\ 1 & -1 & 0 \end{bmatrix} \qquad (2)$$

where $S_0 = (9/8)(d_{111}/d_{11\bar{1}} - 1)$, $d_{111}$ and $d_{11\bar{1}}$ corresponds to 111 and $11\bar{1}$ plane spacings, respectively. For rhombohedral domains, we used $S_0 = 0.63\%$.

For a given condition, the total spontaneous strain matrix of a crystallite is obtained by summation of spontaneous strain matrices of domains corresponding to each of these fourteen different polarization states (six for tetragonal domains and eight for rhombohedral domains), weighted by their respective volume fractions.

Appendix 2: Tensorial transformation for polycrystalline ensemble

A polycrystalline microstructure is simulated by introducing relative orientations between the crystallographic axes (or the LGD frame) and the macroscopic frame that is used to define direction of applied electric-field/stress. To accomplish the above, we define the three Eulerian angles that define the mutual orientation between the two set of axes: (1) one that is defined by the pseudocubic crystallographic axes (or the LGD frame), and (2) macroscopic frame that is convenient to define the applied electric-field/stress or induced polarization/strain response. The three Eulerian angles are defined as follows:

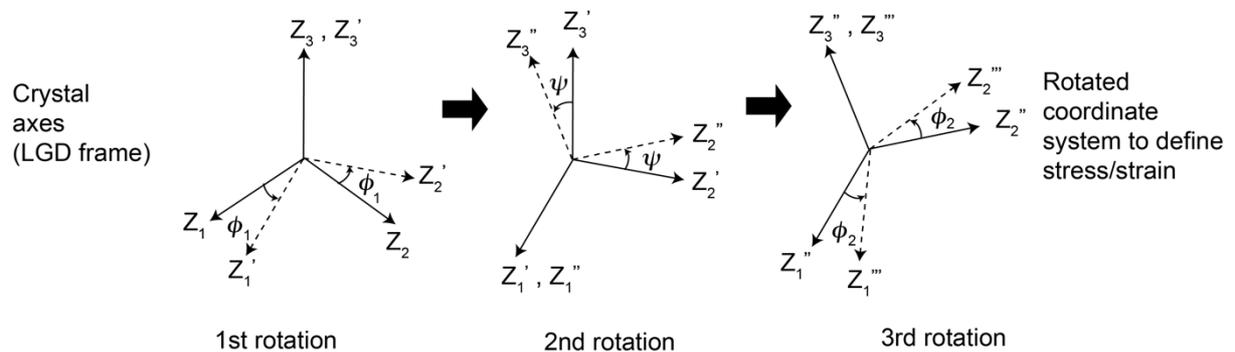

Figure A.2: Definition of the Euler angles that define mutual orientation between the crystal axes (fixed) and the macroscopic sample coordinates.

In the above illustration, the crystal axes are kept fixed while the macroscopic frame is rotated with respect to fixed crystal axes. The first rotation is a counterclockwise rotation around axis $Z_3$ by angle $\psi$, which is defined by the transformation matrix:

$$a_I = \begin{bmatrix} cos\phi_1 & sin\phi_1 & 0 \\ -sin\phi_1 & cos\phi_1 & 0 \\ 0 & 0 & 1 \end{bmatrix}$$

The second rotation is a counterclockwise rotation around new axis $Z_1'$ by angle $\theta$, which is defined by the transformation matrix:

$$a_{II} = \begin{bmatrix} 1 & 0 & 0 \\ 0 & cos\psi & sin\psi \\ 0 & -sin\psi & cos\psi \end{bmatrix}$$

The third rotation is a counterclockwise rotation around new axis $Z_3''$ by angle $\phi$, which is defined by the transformation matrix:

$$a_{III} = \begin{bmatrix} cos\phi_2 & sin\phi_2 & 0 \\ -sin\phi_2 & cos\phi_2 & 0 \\ 0 & 0 & 1 \end{bmatrix}$$

The product of the above three rotations is given by,

$$a = \begin{bmatrix} cos\phi_2 & sin\phi_2 & 0 \\ -sin\phi_2 & cos\phi_2 & 0 \\ 0 & 0 & 1 \end{bmatrix} \begin{bmatrix} 1 & 0 & 0 \\ 0 & cos\psi & sin\psi \\ 0 & -sin\psi & cos\psi \end{bmatrix} \begin{bmatrix} cos\phi_1 & sin\phi_1 & 0 \\ -sin\phi_1 & cos\phi_1 & 0 \\ 0 & 0 & 1 \end{bmatrix}$$

The relationships defining the electric-field(**E**)/polarization(**P**)/stress($\sigma$)/strain($\varepsilon$) state in coordinate system of the crystal axes (LGD frame) and the electric-field(**E**′), polarization(**P**′)/stress($\sigma'$)/strain($\varepsilon'$) state in rotated coordinate system, as depicted in the above figure, are given by:

$$\mathbf{E}' = a\mathbf{E}$$
$$\mathbf{P}' = a\mathbf{P}$$
$$\mathbf{E} = a^T\mathbf{E}'$$
$$\mathbf{P} = a^T\mathbf{P}'$$
$$\sigma' = a\sigma a^T$$
$$\varepsilon' = a\varepsilon a^T$$
$$\sigma = a^T\sigma' a$$
$$\varepsilon = a^T\varepsilon' a$$

The above relationships are reversed if the crystal axes are rotated with respect to a fixed macroscopic frame. For convenience, one can choose whether to fix the orientation of the crystal axes (such as for calculation of inverse pole figures, viz. 8) or to fix the macroscopic frame (such as for calculation of material response, viz. 9,11,12). Both approaches provide identical results.

The inverse pole figure in Figure A.3. depicts the relative orientation between the crystal axes and macroscopic coordinates for a simulated microstructure with 20,000 grains. Each dot represents the angular orientation of the applied electric field with respect to the <100> crystal axes.

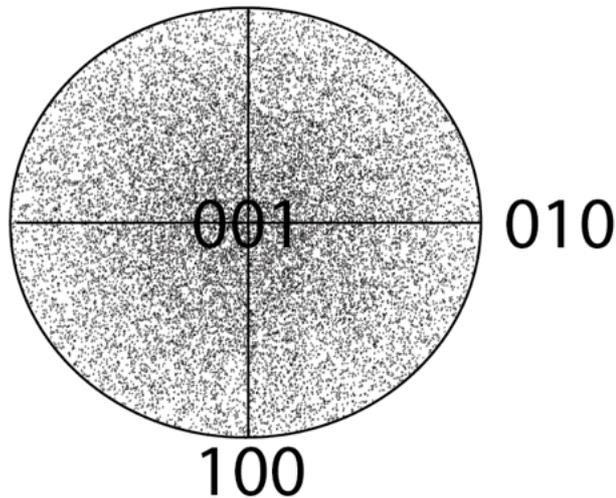

Figure A.3: Inverse pole figure depicting the relative orientation between the crystal axes and orientation of applied electric field.

The simulated random texture is checked for isotropy by computing the polarization responses for different directions of applied electric fields for a polycrystalline material, as shown in the figure A.4 below.

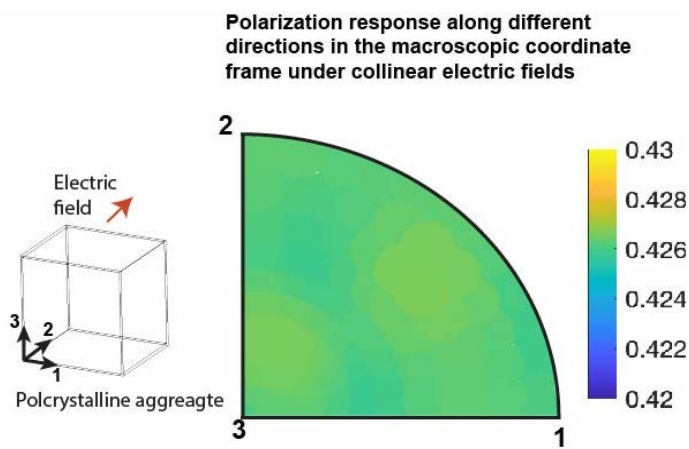

Figure A.4. Polarization response of a polycrystalline material with random texture for different directions of collinear electric fields. The maximum deviation in the overall polarization response is within ±0.5%, which confirms isotropic characteristic of the simulated random texture.

Appendix 3: Material parameters

The value of the Landau coefficients for certain compositions of PZT are given below,[21] which can be used to compute stiffness values for intermediate values of Zr/Ti ratio using regression analysis within definite composition ranges

|  | Mole fraction of PbTiO$_3$ in PZT | | | | | | | | | |
|---|---|---|---|---|---|---|---|---|---|---|
|  | 0.1 | 0.2 | 0.3 | 0.4 | 0.5 | 0.6 | 0.7 | 0.8 | 0.9 | 1.0 |
| P$_s$ (C/m$^2$) | 0.57 | 0.66 | 0.65 | 0.50 | 0.50 | 0.57 | 0.64 | 0.70 | 0.74 | 0.75 |
| T$_C$ (°C) | 256.5 | 300.6 | 334.4 | 364.4 | 392.6 | 418.4 | 440.2 | 459.1 | 477.1 | 492.1 |
| C (10$^5$ °C) | 2.050 | 2.083 | 2.153 | 2.424 | 4.247 | 2.664 | 1.881 | 1.642 | 1.547 | 1.500 |
| $\alpha_{11}$ (10$^7$ m$^5$/C$^2$F) | 41.25 | 31.29 | 22.30 | 13.62 | 4.764 | 3.614 | 0.6458 | -3.050 | -5.845 | -7.253 |
| $\alpha_{12}$ (10$^8$ m$^5$/C$^2$F) | -4.222 | -0.0345 | 1.688 | 2.391 | 1.735 | 3.233 | 5.109 | 6.320 | 7.063 | 7.500 |
| $\alpha_{111}$ (10$^8$ m$^9$/C$^4$F) | 5.068 | 4.288 | 3.560 | 2.713 | 1.336 | 1.859 | 2.348 | 2.475 | 2.518 | 2.606 |
| $\alpha_{112}$ (10$^8$ m$^9$/C$^4$F) | 34.45 | 18.14 | 15.27 | 12.13 | 6.128 | 8.503 | 10.25 | 9.684 | 8.099 | 6.100 |
| $\alpha_{123}$ (10$^8$ m$^9$/C$^4$F) | -8.797 | -7.545 | -7.052 | -5.690 | -2.894 | -4.063 | -5.003 | -4.901 | -4.359 | -3.660 |

* antiferroelectric composition is not included

$$\alpha_1 = (T - T_C)/(2 \times C \times \varepsilon_0)$$

The elastic and electrostrictive coefficients used in the calculations are obtained from ref. [21,51].